\documentclass[useAMS,usenatbib,referee]{mn2e}

\usepackage{amsmath, amssymb}
\usepackage{diagbox}
\usepackage{import}
\usepackage{color}
\usepackage{graphicx}
\usepackage{numprint}

\newcommand{\rot}{\mathbf{\nabla} \times}
\newcommand{\divg}{\mathbf{\nabla}\cdot}

\newcommand{\rlight}{r_{\rm L}}
\newcommand{\Rs}{R_{\rm s}}

\newcommand{\er}{\mathbf{e}_{\rm r}}
\newcommand{\etheta}{\mathbf{e}_\vartheta}
\newcommand{\ephi}{\mathbf{e}_\varphi}

\def\cor#1{{#1}}

\title[GR multipolar electromagnetic fields]{Multipolar electromagnetic fields around neutron stars: general-relativistic vacuum solutions}

\author[J. P\'etri]{J.  P\'etri$^{1}$
\thanks{E-mail: jerome.petri@astro.unistra.fr} \\
  $^{1}$Université de Strasbourg, CNRS, Observatoire astronomique de Strasbourg, UMR 7550, F-67000 Strasbourg, France.}

\begin{document}

\date{Accepted . Received ; in original form }

\pagerange{\pageref{firstpage}--\pageref{lastpage}} 
\pubyear{2016}

\maketitle

\label{firstpage}

\begin{abstract}
Magnetic fields inside and around neutron stars are at the heart of pulsar magnetospheric activity. Strong magnetic fields are responsible for quantum effects, an essential ingredient to produce leptonic pairs and the subsequent broadband radiation. The variety of electromagnetic field topologies could lead to the observed diversity of neutron star classes. Thus it is important to include multipolar components to a presumably dominant dipolar magnetic field. Exact analytical solutions for these multipoles in Newtonian gravity have been computed in recent literature. However, flat spacetime is not adequate to describe physics in the immediate surrounding of neutron stars. We generalize the multipole expressions to the strong gravity regime by using a slowly rotating metric approximation such as the one expected around neutron stars. Approximate formulas for the electromagnetic field including frame dragging are computed from which we estimate the Poynting flux and the braking index. Corrections to leading order in compactness and spin parameter are presented. As far as spindown luminosity is concerned, it is shown that frame dragging remains irrelevant. \cor{For high order multipoles starting from the quadrupole, the electric part can radiate more efficiently than the magnetic part.} Both analytical and numerical tools are employed.
\end{abstract}


\section{Introduction}

Our understanding of pulsar physics has benefited from recent advances in multi-wavelength observational campaigns as well as from developments of new numerical tools able to investigate carefully its magnetosphere and the subsequent radiation mechanisms from a theoretical point of view. As the quality and quantity of data increases regularly, theoreticians are forced to improve the physics of their models in conjunction with the precision of their predictions. Detailed analysis of pulsar phase-resolved spectra and polarisation properties requires accurate models dealing with all possible perturbations departing from a too simple description of pulsar magnetospheres. General relativity belongs to one of these additional physical ingredient compulsory to investigate properly neutron star electrodynamics. This fact became clear in the past years. Indeed efficient pair creation in force-free magnetospheres seems to require frame dragging effects close to the magnetic poles \citep{philippov_ab_2015}. \cite{petri_general-relativistic_2016} investigated in depth general-relativistic force-free magnetospheres. In the same vain \cite{ruiz_pulsar_2014} look at the spindown luminosity and attempted to match neutron star exterior described in the force-free regime to its interior described by relativistic MHD, see also \cite{paschalidis_new_2013}. \cite{konno_general_2000} investigated the impact of general relativistic corrections to curvature radiation. Oscillations of neutron stars in general relativity were also of interest \citep{kojima_approximate_2000}. \cite{morozova_general_2010} studied the influence of neutron star oscillations in general relativity on the plasma density in the magnetosphere for a aligned rotator. Curvature of space time and frame dragging effects on the surrounding electromagnetic field was already emphasized by \cite{beskin_general_1990} and \cite{muslimov_general_1992}.

Any rotating field can be expanded into multipolar components. Thus \cite{bonazzola_general_2015} and \cite{petri_multipolar_2015} showed how to compute exact analytical solutions to multipoles in closed form for flat space-time. Extension to general relativity is highly desirable. The simplest case is a rotating dipole for which \cite{kojima_stationary_2004} gave an approximate solution in general relativity comparing their results with previous analytical work of \cite{rezzolla_general_2001}. \cor{An analytical estimate for the dipole spindown has been given by \cite{rezzolla_electromagnetic_2004}}. Recently \cor{these} authors extended their analysis of oscillations by adding damping due to heating and Joule dissipation \citep{rezzolla_electromagnetic_2016}. In this paper, we show how to extend those results to any multipole and to any order of accuracy.

Maxwell equations in presence of strong gravity using the 3+1 formalism are used to solve for arbitrary electromagnetic field \cor{configurations} in vacuum as presented in Sec.~\ref{sec:Equations}. Exact solutions for static multipolar magnetic fields in Schwarzschild background metric in terms of hypergeometric functions are reminded in Sec.~\ref{sec:StaticMultipole}. In the same gravitational field, exact analytical solutions are found in terms of local confluent Heun functions as explained in Sec.~\ref{sec:RotatingMultipole}. Frame dragging is included in an approximate fashion by numerically solving the system of elliptic equations (Helmholtz equations) for the unknown fields as exposed in Sec.~\ref{sec:ApproximateRotatingMultipole}. \cor{Our analytical treatment is supported by time-dependent numerical simulations of Maxwell equations in general relativity as presented in Sec.~\ref{sec:Simulations}.} Conclusions are drawn in Sec.~\ref{sec:Conclusion}.

\section{Maxwell equations and their solutions}
\label{sec:Equations}

In this section the general formalism to solve Maxwell equations in general-relativistic vacuum is reviewed. For the 3+1~split of spacetime we follow the conventions and definitions given by \cite{petri_general-relativistic_2013}.

\subsection{Split of space-time metric}

We split space-time into a 3+1~foliation such that the background metric~$g_{ik}$ is expressed as
\begin{equation}
  \label{eq:metrique}
  ds^2 = g_{ik} \, dx^i \, dx^k = \alpha^2 \, c^2 \, dt^2 - \gamma_{ab} \, ( dx^a + \beta^a \, c\,dt ) \, (dx^b + \beta^b \, c\,dt )
\end{equation}
where $x^i = (c\,t,x^a)$, $t$ is the time coordinate or universal time and $x^a$ some associated space coordinates. The metric signature is given by $(+,-,-,-)$. $\alpha$ is the lapse function, $\beta^a$ the shift vector and $\gamma_{ab}$ the spatial metric of absolute space. By convention, latin letters from $a$ to $h$ are used for the components of vectors in absolute space, in the range~$\{1,2,3\}$, whereas latin letters starting from $i$ are used for four dimensional vectors and tensors, in the range~$\{0,1,2,3\}$. A fiducial observer (FIDO) is defined by its 4-velocity~$n^i$ such that
\begin{subequations}
 \begin{align}
  n^i & = \frac{dx^i}{d\tau} = \frac{c}{\alpha} \, ( 1, - \bbeta) \\
  n_i & = (\alpha \, c, \textbf{\textit{0}}) .
 \end{align}
\end{subequations}
This vector is orthogonal to the hyper-surface of constant time coordinate~$\varSigma_t$. Its proper time~$\tau$ is measured according to $d\tau = \alpha\,dt$. The relation between the determinants of the space-time metric~$g$ and the pure spatial metric~$\gamma$ is given by $\sqrt{-g} = \alpha \, \sqrt{\gamma}$. For a slowly rotating neutron star, the lapse function is
\begin{equation}
  \label{eq:Lapse}
  \alpha = \sqrt{ 1 - \frac{\Rs}{r} }
\end{equation}
and the shift vector
\begin{subequations}
 \begin{align}
  \label{eq:Shift}
  c \, \bbeta = & - \omega \, r \, \sin\vartheta \, \ephi \\
  \omega = & \frac{\Rs\,a\,c}{r^3} .
 \end{align}
\end{subequations}
We use a spherical coordinate system~$(r,\vartheta,\varphi)$ and an orthonormal spatial basis~$(\er,\etheta,\ephi)$. The metric of a slowly rotating neutron star remains close to the usual flat space, except for the radial direction. Indeed the components of the spatial metric are given in Boyer-Lindquist coordinates by
\begin{equation}
  \label{eq:Metric3D}
  \gamma_{ab} =
  \begin{pmatrix}
    \alpha^{-2} & 0 & 0 \\
    0 & r^2 & 0 \\
    0 & 0 & r^2 \sin^2\vartheta
  \end{pmatrix} .
\end{equation}
For this slow rotation approximation, the spatial metric does not depend on the spin frequency of the massive body but only on its mass~$M$ through $\alpha$. This justifies the slow-rotation approximation. The spin parameter~$a$ is related to the angular momentum~$J$ by $J=M\,a\,c$. It follows that $a$ has units of a length and should satisfy $a \leq \Rs/2$. Introducing the moment of inertia~$I$, we also have $J=I\,\Omega$. In the special case of a homogeneous and uniform neutron star interior with spherical symmetry, the moment of inertia reads
\begin{equation}
  \label{eq:Inertie}
  I = \frac{2}{5} \, M \, R^2 .
\end{equation}
Thus the spin parameter can be expressed as
\begin{equation}
  \label{eq:spin}
  \frac{a}{\Rs} = \frac{2}{5} \, \frac{R}{\Rs} \, \frac{R}{\rlight} .
\end{equation}
For the remainder of this paper, we will use this expression to relate the spin parameter intervening in the metric to the spin frequency of the neutron star. From the above expression, note that the parameter~$a/\Rs$ remains usually small.

\subsection{Maxwell equations in general relativity}

Maxwell equations in 3+1~notation take a traditional form close to the one known in flat space-time except that, as the reader should keep in mind, the three dimensional space is curved and differential operators defined according to the spatial metric~$\gamma_{ab}$. In vacuum, the system reads
\begin{subequations}
\begin{align}
\label{eq:Maxwell1}
 \divg \textbf{\textit{B}} & = 0 \\
\label{eq:Maxwell2}
 \rot \textbf{\textit{E}} & = - \frac{1}{\sqrt{\gamma}} \, \partial_t (\sqrt{\gamma} \, \textbf{\textit{B}}) \\
\label{eq:Maxwell3}
 \divg \textbf{\textit{D}} & = 0 \\
\label{eq:Maxwell4}
 \rot \textbf{\textit{H}} & = \frac{1}{\sqrt{\gamma}} \, \partial_t (\sqrt{\gamma} \, \textbf{\textit{D}}) .
\end{align}
\end{subequations}
The three dimensional vector fields are not independent, they are related by two important constitutive relations, namely
\begin{subequations}
\begin{align}
\label{eq:ConstitutiveE}
  \varepsilon_0 \, \textbf{\textit{E}} & = \alpha \, \textbf{\textit{D}} + \varepsilon_0\,c\,\bbeta \times \textbf{\textit{B}} \\
\label{eq:ConstitutiveH}
  \mu_0 \, \textbf{\textit{H}} & = \alpha \, \textbf{\textit{B}} - \frac{\mathbf\beta \times \textbf{\textit{D}}}{\varepsilon_0\,c}
\end{align}
\end{subequations}
$\varepsilon_0$ is the vacuum permittivity and $\mu_0$ the vacuum permeability. The curvature of absolute space is taken into account by the lapse function factor~$\alpha$ in the first term on the right-hand side and the frame dragging effect is included in the second term, the cross-product between the shift vector~$\bbeta$ and the fields. Space curvature and frame dragging have therefore an imprint on the constitutive relations eq.~(\ref{eq:ConstitutiveE}), (\ref{eq:ConstitutiveH}). From the auxiliary fields $(\textbf{\textit{E}}, \textbf{\textit{H}})$ we get the Poynting flux through a sphere of radius $r$ by computing the two dimensional integral on this sphere by
\begin{equation}
\label{eq:Poynting}
 L = \int_\omega \textbf{\textit{E}} \wedge \textbf{\textit{H}} \, r^2 \, d\omega
\end{equation}
where $d\omega$ is the infinitesimal solid angle and $\omega$ the full sky angle of~$4\,\upi$~sr.
\cor{This integral can be computed analytically in the asymptotic regime of large distances given in the wave zone. The Poynting flux should then be interpreted as the power radiated as seen by a distant observer and not the intrinsic spindown as measured on the neutron star surface. Indeed, due to gravitational redshift, energy is degraded when photons propagate from the surface to the observer and this affects also the measured power in the wave zone. Our spindown luminosity is computed according to the normalization done by this distant observer.}

\subsection{General solution to Maxwell equations}

It is formally possible to give arbitrary solutions to Maxwell equations for divergencelessness electric and magnetic fields in vacuum in a stationary regime. Indeed their expansion into vector spherical harmonics reads, neglecting a possible monopolar $\ell=0$ contribution,
\begin{subequations}
\label{eq:Decomposition_HSV_div_0}
\begin{align}
  \label{eq:Decomposition_HSV_div_0_D}
  \mathbf{D}(r,\vartheta,\varphi,t) = & \sum_{\ell=1}^\infty\sum_{m=-\ell}^\ell \left( \rot [f^{\rm D}_{\ell,m}(r) \, \mathbf{\Phi}_{\ell,m}] + g^{\rm D}_{\ell,m}(r) \, \mathbf{\Phi}_{\ell,m} \right) \, e^{-i\,m\,\Omega\,t} \\
  \label{eq:Decomposition_HSV_div_0_B}
  \mathbf{B}(r,\vartheta,\varphi,t) = & \sum_{\ell=1}^\infty\sum_{m=-\ell}^\ell \left( \rot [f^{\rm B}_{\ell,m}(r) \, \mathbf{\Phi}_{\ell,m}] + g^{\rm B}_{\ell,m}(r) \, \mathbf{\Phi}_{\ell,m} \right) \, e^{-i\,m\,\Omega\,t}.
\end{align}
which correspond to stationary solutions expressed in the frame of a distant observer. The $\mathbf{\Phi}_{\ell,m}$ are vector spherical harmonics defined and introduced in recent works by \cite{petri_general-relativistic_2013}. The functions $g^{\rm D}_{\ell,m}$ and $g^{\rm B}_{\ell,m}$ are related to the function $f^{\rm B}_{\ell,m}$ and $f^{\rm D}_{\ell,m}$ according to Maxwell equations by a linear scaling. Indeed, there exists a simple algebraic relation between $g^{\rm D}_{\ell,m}$ and $f^{\rm B}_{\ell,m}$ on one side, and between $g^{\rm B}_{\ell,m}$ and $f^{\rm D}_{\ell,m}$ on the other side such that
\begin{align}
\label{eq:gDvsfB}
 \alpha \, g^{\rm D}_{\ell,m} = & + i \, \varepsilon_0 \, m \, \tilde{\omega} \, f^{\rm B}_{\ell,m} \\
 \alpha \, g^{\rm B}_{\ell,m} = & - i \,         \mu_0 \, m \, \tilde{\omega} \, f^{\rm D}_{\ell,m} 
\end{align}
\end{subequations}
where $\tilde{\omega} = \Omega - \omega$ is the rotation rate as measured by a local observer. After substitution in Maxwell equations, we get inhomogeneous Helmholtz equations for the potentials $f^{\rm D}_{\ell,m}$ and $f^{\rm B}_{\ell,m}$. Indeed as shown by \cite{petri_general-relativistic_2013}, introducing the Helmholtz operator in curved spacetime by
\begin{equation}
 \label{eq:rdiff}
  \mathcal{W}_{\ell,m}[f] \equiv \alpha^2 \, \left[ \frac{1}{r} \, \frac{\partial}{\partial r} \left( \alpha^2\,\frac{\partial}{\partial r}(r\,f) \right) - \frac{\ell\,(\ell+1)}{r^2} \, f \right] + m^2 \, \frac{\tilde\omega^2}{c^2} \, f 
\end{equation}
the potentials~$f^{\rm D}_{\ell,m}$ must satisfy
\begin{subequations}
\label{eq:Helmholtz}
\begin{align}
\label{eq:HelmholtzD}
 \mathcal{W}_\ell[f^{\rm D}_{\ell,m}] = 3 \, \varepsilon_0 \, \alpha^2 \, \frac{\omega}{r} \, \left[ f^{\rm B}_{\ell-1,m} \, \sqrt{(\ell-1)\,(\ell+1)} \, J_{\ell,m} - f^{\rm B}_{\ell+1,m} \, \sqrt{\ell\,(\ell+2)} \, J_{\ell+1,m} \right]
\end{align}
where $J_{\ell,m}=\sqrt{\frac{\ell^2-m^2}{4\,\ell^2-1}}$ and similarly for the magnetic field~$f^{\rm B}_{\ell,m}$
\begin{align}
\label{eq:HelmholtzB}
 \mathcal{W}_\ell[f^{\rm B}_{\ell,m}] = - 3 \, \mu_0 \, \alpha^2 \, \frac{\omega}{r} \, \left[ f^{\rm D}_{\ell-1,m} \, \sqrt{(\ell-1)\,(\ell+1)} \, J_{\ell,m} - f^{\rm D}_{\ell+1,m} \, \sqrt{\ell\,(\ell+2)} \, J_{\ell+1,m} \right] .
\end{align}
\end{subequations}
The boundary conditions on the neutron star surface are imposed on the electric field in the following way
\begin{equation}
\label{eq:BC}
 \alpha^2 \, \partial_r ( r \, f^D_{\ell,m}) = \varepsilon_0 \, r \, \tilde{\omega} \, \left[ \sqrt{(\ell+1)\,(\ell-1)} \, J_{\ell,m} \, f^B_{\ell-1,m} - \sqrt{\ell\,(\ell+2)} \, J_{\ell+1,m} \, f^B_{\ell+1,m} \right] .
\end{equation}
\cor{This last expression shows that the electric field strength is proportional to $\tilde{\omega}$ which contains the frame dragging effect. This leads to a lowering of the actual rotation rate of the star as seen by a local observer on the surface. Thus frame dragging decreases the electric field intensity and therefore also the spindown luminosity corrections due to rotation of spacetime. However, as shown later in this work, for realistic neutron star parameters, these corrections remain negligible. A second correction is induced by the space curvature and implies a additional factor $\alpha^{-2}$ compared to flat spacetime. The constants of integration are magnified but this effect is sometimes completely cancelled by the general relativistic spherical Hankel functions when $\ell=m$ as shown in the numerical results in Sec.~\ref{sec:ApproximateRotatingMultipole}. For other multipoles with $\ell>m$, compensation is only partial. These conclusions are discussed in detail in the numerical approximate solution Sec.~\ref{sec:ApproximateRotatingMultipole}.}

The Helmholtz operator is conveniently rewritten by introducing the tortoise coordinate $r_*$ such that
\begin{equation}
r_* = r + \Rs \, \ln \left(\frac{r}{\Rs} - 1 \right).
\end{equation}
We are looking for solutions describing outgoing waves that reduce to $e^{i\,k_m\,r}$ in flat space time, thus we introduce another unknown field~$u_{\ell,m}^{\rm B/D}$ such that
\begin{equation}
f_{\ell,m}^{\rm B/D}(r) = u_{\ell,m}^{\rm B/D}(r) \, \frac{e^{i\,k_m\,r_*}}{r}
\end{equation}
where $k_m = m/\rlight$. The curved spacetime Helmholtz operator in terms of these new dependent variables~$u_{\ell,m}$ becomes
\begin{multline}
\mathcal{W}_{\ell,m}[f_{\ell,m}] = \left\{ \frac{1}{r^3} \, \left( 1 - \frac{\Rs}{r} \right) \, \left[ r \, ( r - \Rs ) \, u_{\ell,m}'' + ( 2 \, i \, k_m \, r^2 + \Rs ) \, u_{\ell,m}' - \ell\,(\ell+1) \, u_{\ell,m} \right] \right. \\
 \left. + k_m^2 \, \left[ \left( 1 - \frac{\omega}{\Omega} \right)^2 - 1 \right] \, \frac{u_{\ell,m}}{r} \right\} \, e^{i\,k_m\,r_*} .
\end{multline}
In terms of $u_{\ell,m}^{\rm B/D}$, the inhomogeneous Helmholtz equations become for the electric field
\begin{subequations}
\label{eq:HelmholtzU}
\begin{multline}
 \frac{1}{r} \, \left[ r \, ( r - \Rs ) \, {u^{\rm D}_{\ell,m}}'' + ( 2 \, i \, k_m \, r^2 + \Rs ) \, {u^{\rm D}_{\ell,m}}' - \ell\,(\ell+1) \, u^{\rm D}_{\ell,m} \right] + k_m^2 \, \left[ \left( 1 - \frac{\omega}{\Omega} \right)^2 - 1 \right] \, r \, \frac{u^{\rm D}_{\ell,m}}{\alpha^2} \\ = 3 \, \varepsilon_0 \, \omega \, \left[ u^{\rm B}_{\ell-1,m} \, \sqrt{(\ell-1)\,(\ell+1)} \, J_{\ell,m} - u^{\rm B}_{\ell+1,m} \, \sqrt{\ell\,(\ell+2)} \, J_{\ell+1,m} \right]
\end{multline}
and similarly for the magnetic field
\begin{multline}
 \frac{1}{r} \, \left[ r \, ( r - \Rs ) \, {u^{\rm B}_{\ell,m}}'' + ( 2 \, i \, k_m \, r^2 + \Rs ) \, {u^{\rm B}_{\ell,m}}' - \ell\,(\ell+1) \, u^{\rm B}_{\ell,m} \right] + k_m^2 \, \left[ \left( 1 - \frac{\omega}{\Omega} \right)^2 - 1 \right] \, r \, \frac{u^{\rm B}_{\ell,m}}{\alpha^2} \\
 = - 3 \, \mu_0 \, \omega \, \left[ u^{\rm D}_{\ell-1,m} \, \sqrt{(\ell-1)\,(\ell+1)} \, J_{\ell,m} - u^{\rm D}_{\ell+1,m} \, \sqrt{\ell\,(\ell+2)} \, J_{\ell+1,m} \right] .
\end{multline}
\end{subequations}
For one single multipole with potential specified at the surface by~$f_{\ell,m}^{\rm B}(R)$, the boundary conditions for the electric field reduces to
\begin{subequations}
\label{eq:LimitesfDlm}
\begin{align}
 \alpha^2 \, \partial_r(u_{\ell-1,m}^{\rm D}\,e^{i\,k_m\,r_*}) & = - \varepsilon_0 \, r \, \tilde{\omega} \, \sqrt{(\ell-1)\,(\ell+1)} \, J_{\ell,m} \, f_{\ell,m}^{\rm B} \\
 \alpha^2 \, \partial_r(u_{\ell+1,m}^{\rm D}\,e^{i\,k_m\,r_*}) & = + \varepsilon_0 \, r \, \tilde{\omega} \, \sqrt{\ell\,(\ell+2)} \, J_{\ell+1,m} \, f_{\ell,m}^{\rm B}
\end{align}
\end{subequations}
to be evaluated at \cor{the surface for} $r=R$.

\subsection{Wave zone and Poynting flux}

The Poynting flux of a rotating multipole can be most easily computed in the asymptotic flat spacetime at very large distance. Because of energy conservation law, the flux leaving the star must reach infinity, there is no absorption layer in between. In the wave zone, the expressions~(\ref{eq:Decomposition_HSV_div_0})  can be drastically reduced by the fact that the potential functions behave asymptotically as $f_{\ell,m}(r) \approx u_{\ell,m}^\infty \, e^{i\,\cor{k_m}\,r}/r$ where $\lim_{r\to+\infty} u_{\ell,m}(r) = u_{\ell,m}^\infty$. Neglecting the axisymmetric mode decreasing much faster, like $r^{-(\ell+1)}$, the electromagnetic field becomes in the limit \cor{of large distances} $r\gg\rlight$
\begin{subequations}
\begin{align}
  \label{eq:Solution_Asymptot_B}
  \mathbf{B}_{\rm w} & = \sum_{\ell\geq1,m\neq0} - i \, \frac{e^{i\,m\,(k\,r - \Omega\,t)}}{r}
   \, \cor{k_m \, \left( u^{\rm B,\infty}_{\ell,m} \, \mathbf{\Psi}_{\ell,m} + \mu_0 \, c \, u^{\rm D, \infty}_{\ell,m} \, \mathbf{\Phi}_{\ell,m} \right) } \\
  \label{eq:Solution_Asymptot_D}
  \mathbf{D}_{\rm w} & = \sum_{\ell\geq1,m\neq0} - i \, \frac{e^{i\,m\,(k\,r - \Omega\,t)}}{r} \, \cor{k_m \, \left( u^{\rm D, \infty}_{\ell,m} \, \mathbf{\Psi}_{\ell,m} - \varepsilon_0 \, c \, u^{\rm B, \infty}_{\ell,m} \, \mathbf{\Phi}_{\ell,m} \right) } \\
  \label{eq:OndePlane}
  = & \varepsilon_0 \, c \, \mathbf{B}_{\rm w} \wedge \mathbf{n}.
\end{align}
\end{subequations}
Equation~(\ref{eq:OndePlane}) shows that the solution behaves as a monochromatic plane wave propagating in the radial direction $\mathbf{n} = \mathbf{e}_{\rm r}$ at frequency~$\Omega$. The time averaged Poynting flux is therefore
\begin{equation}
 \mathbf{S} = \frac{\mathbf{D}_{\rm w} \wedge \mathbf{B}_{\rm w}^*}{2\,\mu_0\,\varepsilon_0}
\end{equation}
where $\mathbf{B}_{\rm w}^*$ is the complex conjugate of $\mathbf{B}_{\rm w}$. Integrating the radial component of the Poynting vector along the solid angle~$\omega$ we get the power radiated, using the orthonormality of the vector spherical harmonics, such that
\begin{equation}
\label{eq:luminosite_multipole}
 L = \int_\omega S^{\hat r} \, r^2 \, d\omega = \frac{c}{2\,\mu_0} \, \sum_{\ell \geq 1,m \neq 0} k_m^2 \, \left( |u^{\rm B, \infty}_{\ell,m}|^2 + \mu_0^2 \, c^2 \, |u^{\rm D, \infty}_{\ell,m}|^2 \right) .
\end{equation}
The spin down luminosity~$L$ is independent of the radius as it should from the energy conservation law. Equation~(\ref{eq:luminosite_multipole}) represents the most general expression for the magneto-multipole losses from an arbitrary multipole magnetic field in general relativity. \cor{As soon as the constants of integration~$u^{\rm D/B, \infty}_{\ell,m}$ are known, the full stationary electromagnetic field is determined and its subsequent properties such as spindown and magnetic topology. Our main goal is to fix these constants either with some analytical argument or more accurately via numerical integration of the elliptic problems related to the Helmholtz equations in curved spacetime. We next recall the exact analytical solutions of the static multipoles in general relativity and then look for the stationary rotating multipoles solutions found numerically.}

\section{Exact static multipole solutions}
\label{sec:StaticMultipole}

Finding explicit exact solutions to the rotating multipole problem is difficult because there is no known analytical solution to Helmholtz equations in general relativity in the Schwarzschild metric. Nevertheless, in the static limit of non rotating neutron stars, it is possible to find exact close formula for the multipoles to any order.

From the expansion into vector spherical harmonics eqs.~(\ref{eq:Decomposition_HSV_div_0}), each function $f^{\rm D/\rm B}_{\ell,m}$ has to satisfy a homogeneous second order linear differential equation in Schwarzschild space-time such that
\begin{equation}
\frac{\alpha^2}{r} \, \partial_r (\alpha^2 \, \partial_r (r\,f_{\ell,m})) - \alpha^2 \, \frac{\ell\,(\ell+1)}{r^2} \, f_{\ell,m} = 0 .
\end{equation}
Introducing the new unknown function $\phi_{\ell,m} = r\, f_{\ell,m}$ we get the simple differential equation
\begin{equation}
 \partial_r (\alpha^2 \, \partial_r (\phi_{\ell,m})) - \frac{\ell\,(\ell+1)}{r^2} \, \phi_{\ell,m} = 0
\end{equation}
to be solved with appropriate boundary conditions, namely vanishing potentials at spatial infinity. Moreover, introducing the normalized inverse radial coordinate by $x=\Rs/r$, the functions $\phi_{\ell,m}$ will be solution of 
\begin{equation}
x^2\,(1-x)\,\phi_{\ell,m}'' + x\,(2- 3\,x) \, \phi_{\ell,m}' - \ell\,(\ell+1) \, \phi_{\ell,m} = 0
\end{equation}
which reduces to the hypergeometric differential equation by the change of unknown function $\phi_{\ell,m} = x^\ell\,v_{\ell,m}$ to
\begin{equation}
x\,(1-x)\,v_{\ell,m}'' + (2\,(\ell+1) - ( 2\,\ell+3)\,x) \, v_{\ell,m}' - \ell\,(\ell+2) \, v_{\ell,m} = 0 .
\end{equation}
Setting the parameters $a=\ell, b=\ell+2, c=2\,(\ell+1)$ we indeed find the standard form of the hypergeometric differential equation \citep{olver_nist_2010}. The only solution vanishing at infinity is
\begin{equation}
 \phi_{\ell,m}(x) = C \, x^\ell \, {_2}F_1(\ell,\ell+2,2\,(\ell+1), x)
\end{equation}
which is the expression found by \cite{muslimov_electric_1986}. $C$ is a constant of integration imposed by the boundary condition on the neutron star surface. The solution does not depend on the quantum number~$m$ in the static limit. The \cor{constant~$C$ is chosen such that the} asymptotic value of the potentials converge to their flat spacetime counterpart at large distance. This represents our normalization of the magnetic field strength throughout the paper. We now give the exact analytical expression for the first four multipoles $\ell\in\{1,2,3,4\}$.

\subsection{The magnetic dipole $\ell=1$}

We start our discussion with the general-relativistic magnetic dipole. Introducing the vector spherical harmonics expansion, the static aligned dipole frozen into the neutron star is conveniently written as
\begin{equation}
\label{eq:fB10stat}
 f^{\rm B}_{1,0} = 2 \, \sqrt{6\,\upi} \, \frac{B\,R^3}{\Rs^2} \, \left[ \frac{\ln \left(1-x\right)}{x} + 1 + \frac{x}{2} \right] \cor{ \approx - 2 \, \sqrt{\frac{2\,\upi}{3}} \, \frac{B\,R^3}{r^2} \, \left[ 1 + \frac{3}{4} \, \frac{\Rs}{r} \right]}
\end{equation}
(recall that $x=\Rs/r$). This solution already found by \cite{ginzburg__1964} asymptotes to the flat space-time field at very large distances $r\gg\Rs$. The aligned dipolar magnetic field components are given in an orthonormal basis by
\begin{subequations}
\begin{align}
 B^{\hat r} & = - 3 \, \frac{B\,R^3}{\Rs^3} \, \cos\vartheta \, \mathcal{L}_1(x) \cor{ \approx 2 \, \frac{B\,R^3}{r^3} \, \cos\vartheta \, \left[ 1 + \frac{3}{4} \, \frac{\Rs}{r} \right]} \\
 B^{\hat \vartheta} & = 3 \, \frac{B\,R^3}{\Rs^3 } \, \sin\vartheta \, \mathcal{T}_1(x) \cor{ \approx \frac{B\,R^3}{r^3} \, \sin\vartheta \, \left[ 1 + \frac{\Rs}{r} \right] }\\
 B^{\hat \varphi} & = 0
\end{align}
\end{subequations}
where we introduced the longitudinal and transversal part by
\begin{subequations}
\begin{align}
 \mathcal{L}_1(x) & = x \, (x+2) + 2 \, \ln(1-x) \cor{ \approx - \frac{2}{3} \, x^3 \, \left[ 1 + \frac{3}{4} \, x + o(x^2) \right]} \\
 \mathcal{T}_1(x) & = \frac{(2-x) \, x+2 \, (1-x) \, \ln(1-x)}{\sqrt{1-x}} \cor{ \approx \frac{1}{3} \, x^3 \, \left[ 1 + x + o(x^2) \right] }.
\end{align}
\end{subequations}
The static perpendicular dipole frozen into the neutron star is conveniently written with the normalization
\begin{equation}
\label{eq:fB11stat}
 f^{\rm B}_{1,1} = - \sqrt{2} \, f^{\rm B}_{1,0}
\end{equation}
meaning inclining the previous aligned dipole to $90\degr$ with respect to the rotation axis \cor{leading to the magnetic field components}
\begin{subequations}
\begin{align}
 B^{\hat r} & = - 3 \, \frac{B \, R^3}{\Rs^3} \, e^{i\,\varphi } \, \sin\vartheta \, \mathcal{L}_1(x) \cor{ \approx 2 \, \frac{B\,R^3}{r^3} \, e^{i\,\varphi } \, \sin\vartheta \, \left[ 1 + \frac{3}{4} \, \frac{\Rs}{r} \right] }\\
 B^{\hat \vartheta} & = -3 \, \frac{B\,R^3}{\Rs^3} \, e^{i\,\varphi } \, \cos\vartheta \, \mathcal{T}_1(x) \cor{ \approx - \frac{B\,R^3}{r^3} \, e^{i\,\varphi } \, \cos\vartheta \, \left[ 1 + \frac{\Rs}{r} \right]} \\
 B^{\hat \varphi} & = -3 \, \frac{B\, R^3}{\Rs^3} \, i \, e^{i\,\varphi } \, \mathcal{T}_1(x) \cor{ \approx - \frac{B\,R^3}{r^3} \, i \, e^{i\,\varphi } \, \left[ 1 + \frac{\Rs}{r} \right]} .
\end{align}
\end{subequations} 
\cor{Gravitational corrections to first order expressed by the coefficient $\Rs/r$ are shown to better grab the increase in magnetic field components. The amplification is different depending on the component under consideration thus the field line topology is also modified with respect to a flat spacetime dipole.}

\subsection{The magnetic quadrupole $\ell=2$}

Let us perform the same expansion to the general-relativistic magnetic quadrupole such that it can be expressed inside the star by
\begin{subequations}
\begin{align}
 f^{\rm B}_{2,0} & = - \frac{8}{3} \, \sqrt{\frac{10\,\upi}{3}} \, \frac{B\,R^4}{\Rs^3} \, \frac{x (x (x+6)-24)+6 (3 x-4) \log (1-x)}{x^2} \\
 \cor{ \approx} & \cor{ - 4 \, \sqrt{\frac{2\,\upi}{15}} \, \frac{B\,R^4}{r^3} \, \left[ 1 + \frac{4}{3} \, \frac{\Rs}{r} \right] } .
\end{align}
\end{subequations}
The quadrupolar magnetic field components for the axisymmetric mode $m=0$ are given in an orthonormal basis by
\begin{subequations}
\begin{align}
 B^{\hat r} & = - \frac{10}{3} \, \frac{B\,R^4}{\Rs^4} \, (3 \, \cos 2 \, \vartheta +1) \, \mathcal{L}_2(x) \cor{ \approx \frac{B\,R^4}{r^4} \, (3 \, \cos 2 \, \vartheta + 1 ) \, \left[ 1 + \frac{4}{3} \, \frac{\Rs}{r} \right]} \\
 B^{\hat \vartheta} & = 20 \, \frac{B\,R^4}{\Rs^4} \, \sin 2 \, \vartheta \, \mathcal{T}_2(x) \cor{ \approx 2 \, \frac{B\,R^4}{r^4} \, \sin 2 \, \vartheta \, \left[ 1 + \frac{3}{2} \, \frac{\Rs}{r} \right]} \\
 B^{\hat \varphi} & = 0
\end{align}
\end{subequations}
where we introduced the longitudinal and transversal part by
\begin{subequations}
\begin{align}
 \mathcal{L}_2(x) & = - \frac{x (x (x+6)-24)+6 (3 x-4) \log (1-x)}{x} \\
 & \cor{ = - \frac{3}{10} \, x^4 \, \left[ 1 + \frac{4}{3} \, x + o(x^2) \right] } \\
 \mathcal{T}_2(x) & = \frac{x ((x-12) x+12)+6 (x-2) (x-1) \log (1-x)}{x\,\sqrt{1-x}} \\ 
 & \cor{ = \frac{1}{10} \, x^4 \, \left[ 1 + \frac{3}{2} \, x + o(x^2) \right] } .
\end{align}
\end{subequations}
\cor{The static (2,1) quadrupole frozen into the neutron star is conveniently written with the normalization
\begin{equation}
\label{eq:fB21stat}
 f^{\rm B}_{2,1} = - \frac{1}{2} \, \sqrt{\frac{3}{2}} \, f^{\rm B}_{2,0}
\end{equation}}
\cor{giving} the $m=1$ mode by
\begin{subequations}
\begin{align}
 B^{\hat r} & = -5 \, \frac{B \, R^4}{\Rs^4} \, e^{i \varphi } \, \sin 2\,\vartheta \, \mathcal{L}_2(x) \cor{ \approx \frac{3}{2} \, \frac{B\,R^4}{r^4} \, e^{i \varphi } \, \sin 2\,\vartheta \, \left[ 1 + \frac{4}{3} \, \frac{\Rs}{r} \right] } \\
 B^{\hat \vartheta} & = -10 \, \frac{B \, R^4}{\Rs^4} \, e^{i \varphi } \, \cos 2\,\vartheta \, \mathcal{T}_2(x) \cor{ \approx -\frac{B\,R^4}{r^4} \, e^{i \varphi } \, \cos 2\,\vartheta \, \left[ 1 + \frac{3}{2} \, \frac{\Rs}{r} \right]} \\
 B^{\hat \varphi} & = -10 \, \frac{B \, R^4}{\Rs^4} \, i \, e^{i \varphi } \, \cos \vartheta \,  \mathcal{T}_2(x) \cor{ \approx -\frac{B \, R^4}{r^4} \, i \, e^{i \varphi } \, \cos \vartheta \, \left[ 1 + \frac{3}{2} \, \frac{\Rs}{r} \right] } .
\end{align}
\end{subequations}
\cor{The static (2,2)-quadrupole frozen into the neutron star is conveniently written with the normalization
\begin{equation}
\label{eq:fB22stat}
 f^{\rm B}_{2,2} = \sqrt{\frac{3}{2}} \, f^{\rm B}_{2,0}
\end{equation}}
\cor{leading to} the $m=2$ mode 
\begin{subequations}
\begin{align}
 B^{\hat r} & = -10 \, \frac{B \, R^4}{\Rs^4} \, e^{2 i \varphi } \sin ^2\vartheta \, \mathcal{L}_2(x) \cor{ \approx 3 \, \frac{B\,R^4}{r^4} \, e^{2 i \varphi } \sin ^2\vartheta \, \left[ 1 + \frac{4}{3} \, \frac{\Rs}{r} \right] } \\
 B^{\hat \vartheta} & = - 10 \, \frac{B \, R^4}{\Rs^4} \, e^{2 i \varphi } \sin 2\,\vartheta\, \mathcal{T}_2(x) \cor{ \approx - \frac{B\,R^4}{r^4} \, e^{2 i \varphi } \sin 2\,\vartheta \, \left[ 1 + \frac{3}{2} \, \frac{\Rs}{r} \right] } \\
 B^{\hat \varphi} & = - 20 \, \frac{B \, R^4}{\Rs^4} \, i \, e^{2 i \varphi } \sin \vartheta \, \mathcal{T}_2(x) \cor{ \approx - 2 \, \frac{B \, R^4}{\Rs^4} \, i \, e^{2 i \varphi } \sin \vartheta \,\left[ 1 + \frac{3}{2} \, \frac{\Rs}{r} \right] } .
\end{align}
\end{subequations}

\subsection{The magnetic hexapole $\ell=3$}

Next the magnetic hexapole existing inside the neutron star is written as
\begin{subequations}
\begin{align}
 f^{\rm B}_{3,0} & = \frac{5}{2} \, \sqrt{\frac{7\,\upi}{3}} \, \frac{B\,R^5}{\Rs^4} \, \frac{12 \left(6 x^2-20 x+15\right) \log (1-x)+x (x (x (x+12)-150)+180)}{x^3} \\
 & \cor{ \approx - 2 \, \sqrt{\frac{\upi}{21}} \, \frac{B\,R^5}{r^4} \, \left[ 1 + \frac{15}{8} \, \frac{\Rs}{r} \right]} .
\end{align}
\end{subequations}
The hexapolar magnetic field components for the axisymmetric mode $m=0$ are given in an orthonormal basis by
\begin{subequations}
\begin{align}
B^{\hat r} & = -\frac{35 B \, R^5}{4 \Rs^5} \, \cos \vartheta \, \left(5 \cos ^2\vartheta -3\right) \, \mathcal{L}_3(x) \cor{ \approx \frac{B \, R^5}{r^5} \, \cos \vartheta \, \left(5 \cos^2\vartheta -3\right) \, \left[ 1 + \frac{15}{8} \, \frac{\Rs}{r} \right] } \\
B^{\hat \vartheta} & = \frac{105 B \, R^5}{16 \Rs^5} \,  (\sin \vartheta +5 \sin 3 \, \vartheta ) \, \mathcal{T}_3(x) \cor{ \approx \frac{3}{16} \, \frac{B \, R^5}{r^5} \, \cos \vartheta \, \left(5 \cos ^2\vartheta -3\right) \, \left[ 1 + 2 \, \frac{\Rs}{r} \right] } \\
 B^{\hat \varphi} & = 0
\end{align}
\end{subequations}
where we introduced the longitudinal and transversal part by
\begin{subequations}
\begin{align}
 \mathcal{L}_3(x) & = \frac{\left(12 \left(6 x^2-20 x+15\right) \log (1-x)+x (x (x (x+12)-150)+180)\right)}{x^2} \\
  & \cor{ = - \frac{4}{35} \, x^5 \, \left[ 1 + \frac{15}{8} \, x + o(x^2) \right] }\\
 \mathcal{T}_3(x) & = \frac{(2-x) x ((x-30) x+30)+12 (1-x) ((x-5) x+5) \log (1-x)}{x^2 \, \sqrt{1-x}} \\
  & \cor{ = \frac{1}{35} \, x^5 \, \left[ 1 + 2 \, x + o(x^2) \right] }.
\end{align}
\end{subequations}
\cor{The static (3,1)-quadrupole frozen is normalized according to
\begin{equation}
\label{eq:fB31stat}
 f^{\rm B}_{3,1} = - \frac{32}{\sqrt{3}} \, f^{\rm B}_{3,0}
\end{equation}}
\cor{such that} the $m=1$ mode \cor{becomes}
\begin{subequations}
\begin{align}
 B^{\hat r} & = - 35 \, \frac{B \, R^5}{\Rs^5} \,  e^{i \varphi } \, \left( \sin \vartheta + 5 \, \sin 3\,\vartheta \right) \, \mathcal{L}_3(x) \cor{ \approx 4 \, \frac{B \, R^5}{r^5} \, e^{i \varphi } \, \left( \sin \vartheta + 5 \, \sin 3\,\vartheta \right) \, \left[ 1 + \frac{15}{8} \, \frac{\Rs}{r} \right] } \\
 B^{\hat \vartheta} & = - \frac{35 B \, R^5}{\Rs^5}\, e^{i \varphi } (\cos \vartheta+15 \cos 3\,\vartheta) \, \mathcal{T}_3(x) \cor{ \approx - \frac{B \, R^5}{r^5}\, e^{i \varphi } (\cos \vartheta+15 \cos 3\,\vartheta) \, \left[ 1 + 2 \, \frac{\Rs}{r} \right]} \\
 B^{\hat \varphi} & = - \frac{70 B \, R^5}{\Rs^5}\, i \, e^{i \varphi } (5 \cos 2\,\vartheta+3) \, \mathcal{T}_3(x) \cor{ \approx - 2 \, \frac{B \, R^5}{r^5}\, i \, e^{i \varphi } (5 \cos 2\,\vartheta+3) \, \left[ 1 + 2 \, \frac{\Rs}{r} \right]} .
\end{align}
\end{subequations}
\cor{The static (3,2)-quadrupole frozen is normalized according to
\begin{equation}
\label{eq:fB32stat}
 f^{\rm B}_{3,2} = \sqrt{\frac{2}{15}} \, f^{\rm B}_{3,0}
\end{equation}}
\cor{such that} the $m=2$ mode \cor{gives}
\begin{subequations}
\begin{align}
 B^{\hat r} & = - \frac{35 B \, R^5}{4 \Rs^5} \, e^{2 i \varphi } \sin ^2\vartheta \cos \vartheta \, \mathcal{L}_3(x) \cor{ \approx \frac{B \, R^5}{r^5} \, e^{2 i \varphi } \sin ^2\vartheta \cos \vartheta \, \left[ 1 + \frac{15}{8} \, \frac{\Rs}{r} \right] } \\
 B^{\hat \vartheta} & = \frac{35 \, B \, R^5}{16 \Rs^5} \, e^{2 i \varphi } (\sin \vartheta-3 \sin 3\,\vartheta) \, \mathcal{T}_3(x) \cor{ \approx \frac{B \, R^5}{16 \, r^5} \, e^{2 i \varphi } (\sin \vartheta-3 \sin 3\,\vartheta) \, \left[ 1 + 2 \, \frac{\Rs}{r} \right] } \\
 B^{\hat \varphi} & = - \frac{35  B \, R^5}{4 \Rs^5} \, i \, e^{2 i \varphi } \sin 2\,\vartheta \, \mathcal{T}_3(x) \cor{ \approx - \frac{B \, R^5}{4 \, r^5} \, i \, e^{2 i \varphi } \sin 2\,\vartheta\, \left[ 1 + 2 \, \frac{\Rs}{r} \right] } .
\end{align}
\end{subequations}
\cor{The static (3,3)-quadrupole is normalized to
\begin{equation}
\label{eq:fB33stat}
 f^{\rm B}_{3,3} = \frac{8}{\sqrt{5}} \, f^{\rm B}_{3,0}
\end{equation}}
\cor{thus we get for} the $m=3$ mode 
\begin{subequations}
\begin{align}
 B^{\hat r} & = \frac{35 B \, R^5}{\Rs^5} \, e^{3 i \varphi } \sin ^3\vartheta \, \mathcal{L}_3(x) \cor{ \approx - \frac{4 B \, R^5}{r^5} \, e^{3 i \varphi } \sin ^3\vartheta \, \left[ 1 + \frac{15}{8} \, \frac{\Rs}{r} \right] } \\
 B^{\hat \vartheta} & = \frac{105 B \, R^5}{\Rs^5} \, e^{3 i \varphi } \sin ^2\vartheta \cos \vartheta \, \mathcal{T}_3(x) \cor{ \approx \frac{3 B \, R^5}{r^5} \, e^{3 i \varphi } \sin ^2\vartheta \cos \vartheta\, \left[ 1 + 2 \, \frac{\Rs}{r} \right] } \\
 B^{\hat \varphi} & = \frac{105 B \, R^5}{\Rs^5} \, i \, e^{3 i \varphi } \sin ^2\vartheta \, \mathcal{T}_3(x) \cor{ \approx \frac{3 B \, R^5}{r^5} \, i \, e^{3 i \varphi } \sin ^2\vartheta \, \left[ 1 + 2 \, \frac{\Rs}{r} \right] } .
\end{align}
\end{subequations}

\subsection{The magnetic octopole $\ell=4$}

To finish this discussion about the \cor{lowest} order general-relativistic multipoles, we give the exact solution to the octopole field \cor{for any azimuthal mode~$m$}. Inside the neutron star, the magnetic field is described a priori by a general octopolar expansion such that \cor{the magnetic potential useful for any~$m$ is given with appropriate normalization by}
\begin{subequations}
\begin{align}
 f^{\rm B}_{4,0} & = \frac{336}{5} \, \sqrt{\frac{\upi}{5}} \, \frac{B\,R^6}{\Rs^5 \, x^4} \, [x (x (x (3 x (x+20)-1570)+4620)-3360) + \nonumber \\
 & 60 (5 x (2 (x-6) x+21)-56) \log (1-x) ] \\
 & \cor{ \approx 8 \, \sqrt{\frac{\upi}{5}} \, \frac{B\,R^6}{r^5} \, \left[ 1 + \frac{12}{5} \, \frac{\Rs}{r} \right] }.
\end{align}
\end{subequations}
The octopolar magnetic field components for the axisymmetric mode $m=0$ are given in an orthonormal basis by
\begin{subequations}
\begin{align}
 B^{\hat r} & = - \frac{63 B \, R^6}{20 \Rs^6} \,  \left(9 + 20 \, \cos2\,\vartheta + 35 \, \cos 4 \, \vartheta \right) \, \mathcal{L}_4(x) \\
 & \cor{ \approx - \frac{3}{8} \, \frac{B \, R^6}{r^6} \,  \left(9 + 20 \, \cos2\,\vartheta + 35 \, \cos 4 \, \vartheta \right) \, \left[ 1 + \frac{12}{5} \, \frac{\Rs}{r} \right] } \\
 B^{\hat \vartheta} & = - \frac{63 B \, R^6}{\Rs^6} \, (2 \sin 2 \, \vartheta +7 \sin 4 \, \vartheta ) \, \mathcal{T}_4(x) \cor{ \approx  - \frac{3}{2} \, \frac{B \, R^6}{r^6} \, (2 \sin 2 \, \vartheta +7 \sin 4 \, \vartheta ) \, \left[ 1 + \frac{5}{2}\, \frac{\Rs}{r} \right] } \\
 B^{\hat \varphi} & = 0
\end{align}
\end{subequations}
where we introduced the longitudinal and transversal part by
\begin{subequations}
\begin{align}
 \mathcal{L}_4(x) & = \frac{x (3360-x (x (3 x (x+20)-1570)+4620))-60 (5 x (2 (x-6) x+21)-56) \log (1-x)}{x^3} \\
 & = \frac{5}{42} \, x^6  \, \left[ 1 + \frac{12}{5} \, x + o(x^2) \right] \\
 \mathcal{T}_4(x) & = \frac{x (x (x (x (3 x-190)+1030)-1680)+840)+60 (x-2) (x-1) ((x-7) x+7) \log (1-x)}{x^3 \, \sqrt{1-x}} \\
 & = -\frac{1}{42} \, x^6 \, \left[ 1 + \frac{5}{2} \, x + o(x^2) \right] .
\end{align}
\end{subequations}
\cor{The static (4,1)-quadrupole is conveniently written with the normalization
\begin{equation}
\label{eq:fB41stat}
 f^{\rm B}_{4,1} = \frac{8}{\sqrt{5}} \, f^{\rm B}_{4,0}
\end{equation}}
For the $m=1$ mode we get
\begin{subequations}
\begin{align}
 B^{\hat r} & = \frac{63 B \, R^6}{5 \Rs^6} \, e^{i \varphi } \sin 2\,\vartheta \, ( 1 + 7 \, \cos 2\,\vartheta) \, \mathcal{L}_4(x) \cor{ \approx \frac{3 B \, R^6}{2 r^6} \, e^{i \varphi } \sin 2\,\vartheta \, ( 1 + 7 \, \cos 2\,\vartheta) \, \left[ 1 + \frac{12}{5} \, \frac{\Rs}{r} \right] } \\
 B^{\hat \vartheta} & = \frac{126 B \, R^6}{5 \Rs^6} \, e^{i \varphi } (\cos 2\,\vartheta+7 \cos 4 \vartheta ) \, \mathcal{T}_4(x) \cor{ \approx -3\,\frac{B \, R^6}{5 r^6} \, e^{i \varphi } (\cos 2\,\vartheta+7 \cos 4 \vartheta ) \,\left[ 1 + \frac{5}{2} \, \frac{\Rs}{r} \right] } \\
 B^{\hat \varphi} & = \frac{63 B \, R^6}{5 \Rs^6 } \, i \, e^{i \varphi } (9 \cos \vartheta+7 \cos 3 \vartheta ) \, \mathcal{T}_4(x) \cor{ \approx -3 \frac{B \, R^6}{10 r^6 } \, i \, e^{i \varphi } (9 \cos \vartheta+7 \cos 3 \vartheta ) \, \left[ 1 + \frac{5}{2} \, \frac{\Rs}{r} \right]}.
\end{align}
\end{subequations}
\cor{The static (4,2)-quadrupole is conveniently written with the normalization
\begin{equation}
\label{eq:fB42stat}
 f^{\rm B}_{4,2} = \frac{1}{\sqrt{10}} \, f^{\rm B}_{4,0}
\end{equation}}
For the $m=2$ mode we get
\begin{subequations}
\begin{align}
B^{\hat r} & = - \frac{63 B \, R^6}{20 \Rs^6} \, e^{2 i \varphi } ( 3 + 4 \, \cos2\,\vartheta -7 \, \cos4\,\vartheta ) \, \mathcal{L}_4(x) \\
&\cor{ \approx -3 \, \frac{B \, R^6}{8r^6} \, e^{2 i \varphi } ( 3 + 4 \, \cos2\,\vartheta -7 \, \cos4\,\vartheta )\, \left[ 1 + \frac{12}{5} \, \frac{\Rs}{r} \right] } \\
B^{\hat \vartheta} & = \frac{63 B \, R^6}{5 \Rs^6} \,  e^{2 i \varphi } (2 \sin 2\,\vartheta-7 \sin 4\,\vartheta) \, \mathcal{T}_4(x) \cor{ \approx - 3 \frac{B \, R^6}{10 r^6} \,  e^{2 i \varphi } (2 \sin 2\,\vartheta-7 \sin 4\,\vartheta) \, \left[ 1 + \frac{5}{2} \, \frac{\Rs}{r} \right] } \\
B^{\hat \varphi} & = -\frac{63 B \, R^6}{5 \Rs^6} \, i \, e^{2 i \varphi } (3 \sin \vartheta+7 \sin 3\,\vartheta) \, \mathcal{T}_4(x) \cor{ \approx \frac{3 B \, R^6}{10 r^6} \, i \, e^{2 i \varphi } (3 \sin \vartheta+7 \sin 3\,\vartheta)\, \left[ 1 + \frac{5}{2} \, \frac{\Rs}{r} \right] } .
\end{align}
\end{subequations}
\cor{The static (4,3)-quadrupole is conveniently written with the normalization
\begin{equation}
\label{eq:fB43stat}
 f^{\rm B}_{4,3} = \sqrt{\frac{5}{7}} \, f^{\rm B}_{4,0}
\end{equation}}
For the $m=3$ mode we get
\begin{subequations}
\begin{align}
 B^{\hat r} & = \frac{63}{2} \, \frac{B \, R^6}{\Rs^6} \, e^{\cor{3} i \varphi } ( 2 \sin2 \vartheta - 4 \sin 4\vartheta ) \, \mathcal{L}_4(x) \cor{ \approx \frac{15}{4} \, \frac{B \, R^6}{r^6} \, e^{3 i \varphi } ( 2 \sin2 \vartheta - 4 \sin 4\vartheta ) \, \left[ 1 + \frac{12}{5} \, \frac{\Rs}{r} \right] } \\
 B^{\hat \vartheta} & = 126 \, \frac{B \, R^6}{\Rs^6} \, e^{\cor{3} i \varphi } \, ( \cos2\vartheta - \cos 4\,\vartheta) \, \mathcal{T}_4(x) \cor{ \approx - 3 \, \frac{B \, R^6}{r^6} \, e^{3 i \varphi } \, ( \cos2\vartheta - \cos 4\,\vartheta)\, \left[ 1 + \frac{5}{2} \, \frac{\Rs}{r} \right] } \\
 B^{\hat \varphi} & = 189 \,\frac{B \, R^6}{\Rs^6} \, i \, e^{\cor{3} i \varphi } \, ( \cos \vartheta - \cos3\vartheta) \, \mathcal{T}_4(x) \cor{ \approx - \frac{9}{2} \, \frac{B \, R^6}{r^6} \, i \, e^{3 i \varphi } \, ( \cos \vartheta - \cos3\vartheta)\, \left[ 1 + \frac{5}{2} \, \frac{\Rs}{r} \right] } .
\end{align}
\end{subequations}
\cor{The static (4,3)-quadrupole is conveniently written with the normalization
\begin{equation}
\label{eq:fB44stat}
 f^{\rm B}_{4,4} = \sqrt{\frac{10}{7}} \, f^{\rm B}_{4,0}
\end{equation}}
For the $m=4$ mode we get
\begin{subequations}
\begin{align}
 B^{\hat r} & = - 126  \, \frac{B \, R^6}{\Rs^6} \, e^{\cor{4} i \varphi } \sin^4\vartheta \, \mathcal{L}_4(x) \cor{ \approx  - 15 \, \frac{B \, R^6}{r^6} \, e^{4 i \varphi } \sin^4\vartheta \,\, \left[ 1 + \frac{12}{5} \, \frac{\Rs}{r} \right] } \\
 B^{\hat \vartheta} & = - 504 \, \frac{B \, R^6}{\Rs^6} \, e^{\cor{4} i \varphi } \sin 3\,\vartheta \, \cos\vartheta \, \mathcal{T}_4(x) \cor{ \approx 12 \, \frac{B \, R^6}{r^6} \, e^{4 i \varphi } \sin 3\,\vartheta \, \cos\vartheta \, \left[ 1 + \frac{5}{2} \, \frac{\Rs}{r} \right] } \\
 B^{\hat \varphi} & = - 504 \, \frac{B \, R^6}{\Rs^6} \, i \, e^{\cor{4} i \varphi } \, \sin^3\vartheta \, \mathcal{T}_4(x) \cor{ \approx 12 \, \frac{B \, R^6}{r^6} \, i \, e^{4 i \varphi } \, \sin^3\vartheta \, \left[ 1 + \frac{5}{2} \, \frac{\Rs}{r} \right] } .
\end{align}
\end{subequations}
All these expressions for the first multipoles will be very useful for forthcoming general-relativistic time-dependent simulations of Maxwell equations to look deeper into stationary solutions of the electromagnetic field in vacuum \cor{as well as in force-free magnetospheres}.

\section{Exact rotating multipole fields without frame dragging}
\label{sec:RotatingMultipole}

Frame dragging induces a coupling between different components rendering an analytical solution difficult to find. This is seen in the right-hand side term in eqs.~(\ref{eq:HelmholtzU}) where terms in $\omega$ appear. However, in order to catch the essentials of general-relativistic effects on a rotating multipole, we start with a Schwarzschild background metric, thus neglecting frame-dragging. We justify this approximation a posteriori when solving numerically the system of equations~(\ref{eq:HelmholtzU}), at least for realistic neutron star parameters. This allows a treatment very similar to flat spacetime as shown below.

\subsection{General treatment}

From the expansion into vector spherical harmonics, each function $f^{\rm D/\rm B}_{\ell,m}$ has to satisfy a scalar wave equation in Schwarzschild space-time such that
\begin{equation}
\frac{\alpha^2}{r} \, \partial_r (\alpha^2 \, \partial_r (r\,f)) + \left( k_m^2 - \alpha^2 \, \frac{\ell\,(\ell+1)}{r^2} \right) \, f = 0 .
\end{equation}
Using the tortoise coordinate $r_*$ and the new unknown function $\phi = r\,f$ we arrive at
\begin{equation}
\partial^2_{r_*} \phi + \left( k_m^2 - \alpha^2 \, \frac{\ell\,(\ell+1)}{r^2} \right) \, \phi = 0 .
\end{equation}
We are looking for solutions describing outgoing waves that reduce to $e^{i\,k_m\,r}$ in flat space time, thus we introduce another unknown field~$u$ as
\begin{equation}
\phi = u \, e^{i\,k_m\,r_*} .
\end{equation}
Therefore $u$ is solution of the ordinary differential equation
\begin{equation}
 \alpha^2 \, u'' + \left( 2 \, i \, k_m + \frac{\Rs}{r^2} \right) \, u' - \frac{\ell\,(\ell+1)}{r^2} \, u = 0
\end{equation}
or written more explicitly
\begin{equation}
 r \, (r-\Rs) \, u'' + \left( 2 \, i \, k_m \, r^2 + \Rs \right) \, u' - \ell\,(\ell+1) \, u = 0 .
\end{equation}
Outgoing waves at infinity imposes a finite value for $u$ such that $\lim_{r\rightarrow+\infty} |u(r)| = |u|^\infty< \infty$.
Changing to normalized radius defined by $z=r/\Rs$ and wavenumber by $\epsilon_m = k_m\,\Rs=\Rs/\rlight\ll1$, this equation then becomes
\begin{equation}
\label{eq:HeunConfluent}
 z \, (z-1) \, u''(z) + ( 1 + 2 \, i \, \epsilon_m \, z^2 ) \, u'(z) - \ell\,(\ell+1) \, u(z) = 0 .
\end{equation}
Equation~(\ref{eq:HeunConfluent}) is known as the confluent Heun equation. It has two regular singular points located respectively at the origin of the coordinate system~$z=0$ and at the Schwarzschild radius~$z=1$ and one irregular singular point at spatial infinity~$z=+\infty$ \citep{olver_nist_2010}. In the standard non symmetrical canonical form given by \cite{ronveaux_heuns_1995}, the confluent Heun equation is usually summarized by
\begin{equation}
\frac{d^2 w}{dz^2} + \left( 4\,p + \frac{\gamma}{z} + \frac{\delta}{z-1} \right) \, \frac{dw}{dz} + \frac{4\,p\,\alpha\,z-\sigma}{z\,(z-1)} \, w = 0 .
\end{equation}
In the case of eq~(\ref{eq:HeunConfluent}), the parameters are defined by
\begin{subequations}
\begin{align}
 4\,p & = 2 \, i \, \epsilon_m \\
 \gamma & = -1 \\
 \delta & = 1 + 2 \, i \, \epsilon_m \\
 \alpha & = 0 \\
 \sigma & = \ell\,(\ell+1) .
\end{align}
\end{subequations}
Local solutions to the confluent Heun equation around the regular singular points $z=0,1$ can be expressed as a Frobenius series expansion. Unfortunately, such expansions do not extend to $z=+\infty$ because their radius of converge is at most equal to the distance to the next singular point, thus a convergence around the point~$z_0$ only within $|z-z_0|<1$. The solution called radial \cor{function} and denoted by ${\rm Hc}^{(\rm r)}(p,\alpha,\gamma,\delta,\sigma;z)$ remaining finite at infinite radius \citep{cook_gravitational_2014} is defined by 
\begin{equation}
 \lim_{z\rightarrow+\infty} z^\alpha \, {\rm Hc}^{(\rm r)}(p,\alpha,\gamma,\delta,\sigma;z) = 1 .
\end{equation}
The two independent local solutions at infinity are therefore
\begin{subequations}
\begin{align}
& Hc^{(\rm r)}(p,\alpha,\gamma,\delta,\sigma;z) \\
e^{-4\,p\,z} \, & Hc^{(\rm r)}(-p,-\alpha+\gamma+\delta,\gamma,\delta,\sigma-4\,p\,\gamma;z) .
\end{align}
\end{subequations}
In our case, $\alpha=0$ therefore $\lim_{z\rightarrow+\infty} {\rm Hc}^{(\rm r)}(p,\alpha,\gamma,\delta,\sigma;z) = 1$ which is indeed compatible with the boundary condition \cor{we impose on $u_{\ell,m}$. These constants should remain finite and} different from zero. The other solution behaves like $e^{-4\,p\,z} \, z^{\alpha-\gamma-\delta} = e^{-2 \, i \, \epsilon_m \, (z+\ln z) } = e^{-2 \, i \, k_m \, (r + \Rs\,\ln (r/\Rs)) }$ which is the solution for ingoing wave from infinity that we discard in our analysis. The asymptotic series at infinity given as Thom\'e solution for the function ${\rm Hc}^{(\rm r)}(p,\alpha,\gamma,\delta,\sigma;z)$ is \citep{philipp_analytic_2015}
\begin{equation}
\label{eq:SeriesHc}
{\rm Hc}^{(\rm r)}(p,\alpha,\gamma,\delta,\sigma;z) = \sum_{n=0}^{+\infty} a_n \, z^{-n}
\end{equation}
where the coefficients $a_n$ satisfy a three-term recurrence
\begin{subequations}
\begin{align}
f_n & \, a_{n+1} + g_n \, a_n + h_n \, a_{n-1} = 0 \\
f_n & =- 2 \, i \, k \, \Rs \, (n+1) \\
g_n & = n \, ( n + 1 ) - \ell \, ( \ell + 1 ) \\
h_n & = - (n-1) \, (n+1)
\end{align}
\end{subequations}
leading unfortunately to the divergence of the series \cor{proposed in} eq.~(\ref{eq:SeriesHc}) \citep{ronveaux_heuns_1995}. 

In Newtonian gravity, $\Rs=0$ and the three-term recurrence simplifies into a two-term recurrence because $f_n=0$ and it is finite, stopping whenever $n=\ell$. The solutions are polynomials in $z^{-1}$ and corresponds to the \cor{standard} spherical Hankel functions $h_\ell^{(1)}(k_m\,r)$ \cor{as used by} \cite{petri_multipolar_2015}. In the general-relativistic static limit, solutions are found by setting $\epsilon_m=0$ in eq.~(\ref{eq:HeunConfluent}) which means $k_m=0$ and therefore no rotation. In that case, the confluent Heun equation reduces to the hypergeometric differential equation of Sec.~\ref{sec:StaticMultipole}.

Series solutions of the confluent Heun equation~(\ref{eq:HeunConfluent}) are found with Frobenius method and a judicious change of the independent variable \citep{leaver_solutions_1986}. Because of the two regular singular points located respectively at $z=0$ and $z=1$ and the irregular singular point at $z=+\infty$ we use a Jaff\'e transform to expand the solution around the regular singular point $z=1$. According to Jaff\'e, the transform reads
\begin{equation}
 x = \frac{z-1}{z} .
\end{equation}
In that way the regular singular point $z=0$ is rejected at $x=-\infty$, the regular singular point $z=1$ at a value $x=0$ and the irregular singular point $z=+\infty$ at $x=1$. The differential equation then reads
\begin{equation}
x \, (1-x)^2 \, u''(x) + ( 2 \, i \, \epsilon_m + (1-x)\,(1-3\,x) ) \, u'(x) - \ell\,(\ell+1) \, u(x) = 0 .
\end{equation}
This technique was used by \cite{kearney_schwarzschild_1978} to solve electrodynamics problems in Schwarzschild metric applied to black holes and neutron stars. We look for series solution around $x=0$ such that $u(x) = \sum_{n=0}^{+\infty} a_n \, x^n$. We know that the series is convergent up to the nearest singular point thus up to $x=1$ which corresponds to $z=+\infty$. The solution is therefore valid in whole space outside the Schwarzschild radius. Equating the monomes of same order $x^n$ we arrive at the following recurrence relations
\begin{subequations}
\begin{align}
 \ell\,(\ell+1) \, a_0 & = ( 1 + 2 \, i \, \epsilon_m ) \, a_1 \\
 (n+1)\,(n+1+2 \, i \, \epsilon_m)\,a_{n+1} & = [2\,n\,(n+1)\,+\ell\,(\ell+1)]\,a_n - (n+1)\,(n-1)\,a_{n-1} .
\end{align}
\end{subequations}
For very large $n$ with $n\gg1$ the relation reduces to $a_{n+1} = 2\,a_n - a_{n-1}$ therefore the ratio $a_{n+1}/a_n = 2 - a_{n-1}/a_n$ leading to $\lim_{n\rightarrow+\infty} a_{n+1}/a_n=1$. Thus the radius of convergence of the series is indeed $\rho=1$. To find a solution to our problem, we need to check that the series is also convergent at the point $x=1$ that is on the circle of convergence. There is no general theorem on convergence on this circle and the problem is a delicate mathematical question with no definite answer. Actually the series does not converge at $x=1$. It is thus impossible to satisfy the boundary condition at infinity with this series expansion. We leave the solution with the notation ${\rm Hc}^{(\rm r)}(p,\alpha,\gamma,\delta,\sigma;r/\Rs)$ and do not give explicit expressions for it.

Close to the neutron star we have $\epsilon \, z^2 \ll 1$ so that we can neglect this term in front of $u'$. The solution therefore reduces to the general relativistic static multipole as described in \cor{Sec.~\ref{sec:StaticMultipole}.} Let us write $\mathcal{H}_\ell^{(1)}(k_m\,r)$ the solution to the confluent Heun equation that remains finite at infinity. The solution corresponding to a spherical outgoing wave, that with $m>0$, is
\begin{equation}
 u_{\ell,m}^{\rm D/B} = a_{\ell,m}^{\rm D/B} \, {\rm Hc}^{(\rm r)}(p,\alpha,\gamma,\delta,\sigma;z) .
\end{equation}
The constant of integration $a_{\ell,m}^{\rm B}$ is determined from the boundary conditions on the neutron star surface noting that 
\begin{equation}
\label{eq:LimiteuBlm}
 u_{\ell m}^{\rm B}(R) = R \, f_{\ell m}^{\rm B}(R) \, e^{-i\,k_m\,r_*(R)} .
\end{equation}
Moreover we introduced the new function
\begin{equation}
\mathcal{H}_\ell^{(1)}(\cor{z}) = \frac{{\rm Hc}^{(\rm r)}(p,\alpha,\gamma,\delta,\sigma;\cor{z})}{\cor{z}} \, e^{i\,k_m\,r_*}
\end{equation}
as a generalization of the spherical Hankel functions to the \cor{curved metric of Schwarzschild type}. The constant of integration for the magnetic part is therefore
\begin{equation}
\label{eq:aBlm}
 a^{\rm B}_{\ell,m} = \frac{f^{\rm B}_{\ell,m}(R)}{\mathcal{H}_\ell^{(1)}(k_m\,R)} .
\end{equation}
The constant of integration for the electric part obtained from the continuity of the tangential component of the electric field gives 
\begin{equation}
 \cor{\alpha^2} \, a^{\rm D}_{\ell,m} \, \left.\partial_r ( r \, \mathcal{H}_\ell^{(1)}(k_m\,r))\right|_{r=R} = \varepsilon_0 \, R \, \Omega \, \left[ \sqrt{(\ell+1)\,(\ell-1)} \, J_{\ell,m} \, f^{\rm B}_{\ell-1,m}(R) - \sqrt{\ell\,(\ell+2)} \, J_{\ell+1,m} \, f^{\rm B}_{\ell+1,m}(R) \right] .
\end{equation}
From this definition the electric and magnetic potentials are deduced according to
\begin{equation}
 f^{\rm D/B}_{\ell,m}(r) = a^{\rm D/B}_{\ell,m} \, \mathcal{H}_\ell^{(1)}(k_m\,r) .
\end{equation}


\subsection{Solution for one multipole}

Useful exact solutions are given for a particular multipole fields with fixed numbers $(\ell,m)$. Let us assume that inside the star, the magnetic field is solely represented by the function $f^{\rm B}_{\ell,m}(r)$. Then the only non vanishing magnetic field coefficient is given by equation~(\ref{eq:aBlm}) (we discard axisymmetric cases $m=0$ which represent static solutions). Moreover the two non-vanishing electric field coefficients are (if $\ell=1$ only one solution exists, see the dipole case below) for $m>0$
\begin{subequations}
\label{eq:aDlm}
\begin{align}
 \cor{\alpha^2} \, a^{\rm D}_{\ell+1,m} \, \left.\partial_r ( r \, \mathcal{H}_{\ell+1}^{(1)}(k_m\,r))\right|_{r=R} & = \varepsilon_0 \, R \, \Omega \,  \sqrt{\ell\,(\ell+2)} \, J_{\ell+1,m} \, f^{\rm B}_{\ell,m}(R) \\
 \cor{\alpha^2} \, a^{\rm D}_{\ell-1,m} \, \left.\partial_r ( r \, \mathcal{H}_{\ell-1}^{(1)}(k_m\,r))\right|_{r=R} & = - \varepsilon_0 \, R \, \Omega \, \sqrt{(\ell-1)\,(\ell+1)} \, J_{\ell,m} \, f^{\rm B}_{\ell,m}(R) .
 \end{align}
\end{subequations}
We conclude that the solution is fully specified by the three constants of integration $(a^{\rm B}_{\ell,m}, a^{\rm D}_{\ell+1,m}, a^{\rm D}_{\ell-1,m})$ as already demonstrated  in flat spacetime. The Poynting flux associated to this particular solution is for the relevant cases~$m>0$ (it vanishes for axisymmetric cases $m=0$)
\begin{subequations}
\label{eq:luminosite_un_multipole}
\begin{align}
 P_{\ell,m} & = \frac{c}{2\,\mu_0} \, \left[ |a^{\rm B}_{\ell,m}|^2 + \mu_0^2 \, c^2 \, ( |a^{\rm D}_{\ell-1,m}|^2 + |a^{\rm D}_{\ell+1,m}|^2 ) \right] \\
 & = \frac{c|f^{\rm B}_{\ell,m}(R)|^2}{2\,\mu_0} \, \mathcal{S}_{\ell,m} \\
 \mathcal{S}_{\ell,m} & = \frac{1}{|\mathcal{H}_\ell^{(1)}(k_m\,R)|^2} + \frac{R^2}{\rlight^2} \, \left( \frac{\ell(\ell+2)\,J^2_{\ell+1,m}}{|\partial_r ( r \, \mathcal{H}_{\ell+1}^{(1)}(k_m\,r))|^2_R} + \frac{(\ell-1)(\ell+1)\,J^2_{\ell,m}}{|\partial_r ( r \, \mathcal{H}_{\ell-1}^{(1)}(k_m\,r))|^2_R} \right) .
 \end{align}
\end{subequations}
For the special case $\ell=1$, the constant $a^{\rm D}_{\ell-1,m}$ does not exist.

\cor{Expression~(\ref{eq:luminosite_un_multipole}) generalizes the Newtonian formula found in \cite{petri_multipolar_2015}. At this point, it is opportune to stress that the Poynting flux arises mainly not from the single magnetic multipole $(\ell,m)$ depicted by the constant $a^{\rm B}_{\ell,m}$ but from the lowest order multipole~$\ell'$, let it be magnetic or electric. Because of the boundary conditions imposed on the neutron star surface where electric multipoles of order $\ell\pm1$ are induced from the magnetic multipole of order $(\ell,m)$, there are cases where the electric multipole is dominant compared to the magnetic multipole $(\ell,m)$. For the magnetic dipole with quantum numbers $(\ell,m)=(1,1)$, the only electric multipole has quantum numbers $(\ell',m)=(2,1)$ thus represents a higher order multipole and therefore remains negligible to first approximation in spin rate. The Poynting flux $P_{\ell,m}$ is mainly attributed to the constant of integration~$a^{\rm B}_{1,1}$ that is the magnetic dipole.}

\cor{Any electric multipole must satisfy the requirement $m \leqslant \ell' = \ell \pm 1$. Moreover electric multipoles are dominant if their quantum number satisfies~$\ell'<\ell$, thus the only choice is $\ell'=\ell-1<\ell$ and this requires $\ell \geqslant m+1$. We conclude that for single magnetic multipoles given by $(\ell\geqslant m+1,m)$ the electric multipole of order $(\ell-1,m)$ dominates the spindown rate. Let us prove this fact in the flat spacetime geometry where exact analytical formulas for any ratio $R/\rlight$ have been found. Table \ref{tab:SpindownContribution} summarizes the weights associated to each multipole and the percentage of luminosity attributed to the electric multipole (last column). Note that the normalized total luminosity is split into contributions from different multipoles according to $\mathcal{S}_{\ell,m} = \mathcal{C}_{\ell,m} + \mathcal{C}_{\ell-1,m} + \mathcal{C}_{\ell+1,m}$. $\mathcal{C}_{\ell,m}$ represents the contribution from the magnetic multipole through the constant $a^{\rm B}_{\ell,m}$, $\mathcal{C}_{\ell-1,m}$ represents the contribution from the electric multipole through the constant $a^{\rm D}_{\ell-1,m}$ and $\mathcal{C}_{\ell+1,m}$ represents the contribution from the electric multipole through the constant $a^{\rm D}_{\ell+1,m}$. This table emphasizes three points. First, the electric multipole $(\ell+1,m)$ never contributes to the spindown in the the limit of a point multipole. Second, the magnetic multipole $(\ell=m,m)$ is the only relevant contributor to the spindown. Third, for any multipole $(\ell>m,m)$ the contribution from the electric multipole $(\ell-1,m)$ can be substantial and is always larger than the luminosity emanating from the magnetic multipole $(\ell,m)$. Consequently, apart for the magnetic dipole, in order to estimate the spindown luminosity of a single magnetic multipole of order $\ell\geqslant2$, we need to take care of the associated electric multipole to properly deduce the energy loss rate. The discussion focused on the Newtonian case but the same obviously applies to a general-relativistic magnetic multipole as will be shown from the numerical solutions.
\begin{table}
\centering
\begin{tabular}{ccccc}
\hline
Multipole $(\ell,m)$ & $\mathcal{C}_{\ell,m}$ & $\mathcal{C}_{\ell-1,m}$ & $\mathcal{C}_{\ell+1,m}$ & $\mathcal{C}_{\ell-1,m}/\mathcal{S}_{\ell,m}$ \\
\hline
\hline
$(1,1)$ & 1 & - & 0 & 0 \\
\hline
$(2,1)$ & $\dfrac{1}{9}$ & $\dfrac{3}{5}$ & 0 & $\dfrac{27}{32}\approx84$\% \\
$(2,2)$ & $\dfrac{64}{9}$ & - & 0 & 0 \\
\hline
$(3,1)$ & $\dfrac{1}{225}$ & $\dfrac{16}{315}$ & 0 & $\dfrac{80}{87}\approx92$\% \\
$(3,2)$ & $\dfrac{256}{225}$ & $\dfrac{138}{63}$ & 0 & $\dfrac{25}{39}\approx64$\% \\
$(3,3)$ & $\dfrac{729}{25}$ & - & 0 & 0 \\
\hline
$(4,1)$ & $\dfrac{1}{\numprint{11025}}$ & $\dfrac{1}{567}$ & 0 & $\dfrac{175}{184}\approx95$\% \\
$(4,2)$ & $\dfrac{\numprint{1024}}{\numprint{11025}}$ & $\dfrac{\numprint{1024}}{\numprint{2835}}$ & 0 & $\dfrac{35}{44}\approx80$\% \\
$(4,3)$ & $\dfrac{\numprint{6561}}{\numprint{1225}}$ & $\dfrac{27}{5}$ & 0 & $\dfrac{245}{488}\approx50$\% \\
$(4,4)$ & $\dfrac{\numprint{1048576}}{\numprint{11025}}$ & - & 0 & 0 \\
\hline
\end{tabular} 
\caption{Spindown contribution from the magnetic and electric multipoles in the limit of point multipoles ($R/\rlight\rightarrow0$). The last column gives the percentage of spindown luminosity attributed to the dominant electric multipole.}
\label{tab:SpindownContribution}
\end{table}
}

\cor{This apparently misleading results arises because although a factor $(R/\rlight)^2$ appears in front of the electric multipoles, it is compensated by the lower order of one of these electric multipoles containing a spherical Hankel function behaving like $|\partial_r ( r \, \mathcal{H}_{\ell-1}^{(1)}(k_m\,r))|^{-2}_R$. As a result, $(R/\rlight)^2 \, |\partial_r ( r \, \mathcal{H}_{\ell-1}^{(1)}(k_m\,r))|^{-2}_R$ is of the same order in $R/\rlight$ as the magnetic multipole $|\mathcal{H}_\ell^{(1)}(k_m\,R)|^{-2}$ for $R/\rlight\ll1$. Actually, we showed that contrary to being negligible, this electric multipole is dominant for the energy loss rate as soon as $\ell>m$. Indeed, in flat spacetime, the spherical Hankel functions can be expanded to lowest order when $k_m\,R\ll1$. In that case the contributions to spindown luminosities are
\begin{equation}
 \mathcal{S}_{\ell,m} = m^{2\,\ell} \, (k\,R)^{2\,\ell+2} \, \left[ \frac{m^2}{((2\,\ell-1)!!)^2} + \frac{(\ell+1)\,J_{\ell,m}^2}{(\ell-1)\,((2\,\ell-3)!!)^2} \right]
\end{equation}
where the first term accounts for the magnetic multipole of order $(\ell,m)$ and the second term accounts for the electric multipole of order $(\ell-1,m)$ (for $\ell>1$). We neglect terms of higher order in the product $k\,R$, especially those arising from the $(\ell+1,m)$ electric multipole. The term in bracket in front of $\frac{R^2}{\rlight^2}$ in eq.~(\ref{eq:luminosite_un_multipole}) is not necessarily a higher order correction of the spindown in the point multipole limit compared to the first term in $\mathcal{S}_{\ell,m}$. It is actually exactly of the same order as the associated magnetic multipole. 
}

\cor{There exist no simple general-relativistic expression for the spherical Hankel functions $\mathcal{H}_\ell^{(1)}(k_m\,R)$. Nevertheless, following \cite{rezzolla_electromagnetic_2004}, we could adopt the flat spacetime counterpart as a good approximation taking into account the gravitational redshift of the spin frequency at the surface by noting $\Omega_R=\Omega/\alpha_R$ thus replacing $\mathcal{H}_\ell^{(1)}(k_m\,R) = h_\ell^{(1)}(m\,\Omega_R\,R/c)$. Using the asymptotic expression for $h_\ell^{(1)}(x)$ for $x\ll1$, the major contribution to the spindown luminosity should be
\begin{equation}
P_{\ell,m} = \frac{c|f^{\rm B}_{\ell,m}(R)|^2}{2\,\mu_0} \, \frac{1}{(2\,(\ell-1)!!)^2} \, \left( \frac{m\,\Omega\,R}{\alpha_R\,c}\right)^{2\,\ell+2} .
\end{equation}
Compared to flat spacetime, we note two correcting factors. The first one arising from the amplification of the magnetic field strength at the stellar surface through the factor $|f^{\rm B}_{\ell,m}(R,\Rs)/f^{\rm B}_{\ell,m}(R,\Rs=0)|^2$ and the second from the rotation frequency shift through a factor $\alpha_R^{-(2\,\ell+2)}$. These are the results already discussed by \cite{rezzolla_electromagnetic_2004}. The general relativistic luminosity increase is therefore completely determined by the Schwarzschild radius $\Rs$ of the star. There are no corrections including $\rlight$. Unfortunately as we show in this paper and as was already found in \cite{petri_general-relativistic_2013} by time-dependent numerical simulations, this estimate is not correct. For extremely slow rotation rate, when $\rlight\rightarrow+\infty$, the general-relativistic spindown reduces to the Newtonian expression, their ratio tending to one. This is easily understood by the fact that radiation starts at the light cylinder and if this surface is rejected to large distances, with our normalization, the magnetic dipole geometry looks very similar to flat spacetime with no gravitational perturbations.
}

\cor{To support our conclusion on firm basis, we show very accurate numerical solutions of spherical Hankel functions in general relativity up to $\ell=4$. They are thoroughly investigated in the next subsection.}

\subsection{General-relativistic spherical Hankel functions}

\cor{The wave equations in the Schwarzschild metric and in the slow rotation metric are solved numerically to high accuracy with an expansion onto rational Chebyshev polynomials. Here we give some useful numerical approximations for these general-relativistic spherical Hankel functions for different compactness $R/\Rs$ and spin rate $R/\rlight$. The results of our integration of the boundary value problems are given by the functions shown in fig. \ref{fig:ComparaisonHankel1} for the mode $\ell=1$, in fig.\ref{fig:ComparaisonHankel2} for the mode $\ell=2$, in fig.\ref{fig:ComparaisonHankel3} for the mode $\ell=3$, in fig.\ref{fig:ComparaisonHankel4} for the mode $\ell=4$. To ease comparison, we use the same boundary conditions on both end of the integration interval. The solution must vanish at infinity at matches the flat spherical Hankel functions on the left boundary, $\mathcal{H}_{\ell}^{(1)}(R/\rlight) =h_{\ell}^{(1)}(R/\rlight) $.}

\cor{With these same boundary conditions the curved spherical Hankel functions suffer from an amplitude decrease for large distances compared to their flat spacetime equivalents. The decrease gets stronger for larger compactness. This means that taking the flat function as an exact solution to the curved spacetime problem but adapting the inner boundary conditions according to general-relativistic corrections overestimates the Poynting flux as measured by a distant observer. A correct answer requires a careful integration of the curved spacetime wave equation as presented in this paper.}

\cor{The evaluation of the spindown luminosity requires the value of the functions $u^{\rm D/B}$ at infinity and denoted by $u^{\rm D/B}_\infty$. As a comparison between flat and curved spacetime we shown in table~\ref{tab:LimitHankel} these values for the same parameters $R/\Rs$ and $R/\rlight$. These values are useful to compute the energy loss from a magnetic multipole of order $(\ell=m,m)$ but not for the other multipoles. The reason has been exposed previously.
}

\cor{We also integrated the wave equation with frame dragging and did not found significant deviation from Schwarzschild solutions as long as $R/\rlight\ll1$. The solution in Schwarzschild spacetime remain valid to very good accuracy for a slowly rotating neutron star metric. They are not represented here as they overlap to previous plots.}

\begin{table}
\centering
\begin{tabular}{cccc}
\hline
$\ell$ & $\Rs/R=0$ & $\Rs/R=0.5$ (schw) & $\Rs/R=0.5$ (srns) \\
\hline
\hline
1 & -1 & -0.656667  + 0.0962066*i & -0.655596  + 0.0961236*i \\
2 & +i & +0.0821953 + 0.450445*i  & +0.0821234 + 0.449962*i \\
3 & +1 & +0.318341  - 0.0644468*i & +0.318087  - 0.0644009*i \\
4 & -i & -0.0493130 - 0.227628*i  & -0.0492839 - 0.227482*i \\
\hline
\end{tabular}
\caption{Limit of the spherical Hankel functions for several multipoles~$\ell$. The table gives the values of $u^{\rm B,\infty}_{\ell,1}$. $\Rs/R=0$ represents flat spacetime, $\Rs/R=0.5$ (schw) Schwarzschild metric and $\Rs/R=0.5$ (srns) the slowly rotating neutron star metric.}
\label{tab:LimitHankel}
\end{table}

\begin{figure}
\centering
\input{comparaison_hankel_lm11_ri10_NR30_g0.tex}
\caption{Comparison of the spherical Hankel function for $\ell=1$ in flat and Schwarzschild spacetime for $R/\rlight=0.1$ and several ratio~$\Rs/R$ as shown in the legend. Real and imaginary part are shown separately.}
\label{fig:ComparaisonHankel1}
\end{figure}

\begin{figure}
\centering
\input{comparaison_hankel_lm21_ri10_NR30_g0.tex}
\caption{Comparison of the spherical Hankel function for $\ell=2$ in flat and Schwarzschild spacetime for $R/\rlight=0.1$ and several ratio~$\Rs/R$ as shown in the legend. Real and imaginary part are shown separately.}
\label{fig:ComparaisonHankel2}
\end{figure}

\begin{figure}
\centering
\input{comparaison_hankel_lm31_ri10_NR30_g0.tex}
\caption{Comparison of the spherical Hankel function for $\ell=3$ in flat and Schwarzschild spacetime for $R/\rlight=0.1$ and several ratio~$\Rs/R$ as shown in the legend. Real and imaginary part are shown separately.}
\label{fig:ComparaisonHankel3}
\end{figure}

\begin{figure}
\centering
\input{comparaison_hankel_lm41_ri10_NR30_g0.tex}
\caption{Comparison of the spherical Hankel function for $\ell=4$ in flat and Schwarzschild spacetime for $R/\rlight=0.1$ and several ratio~$\Rs/R$ as shown in the legend. Real and imaginary part are shown separately.}
\label{fig:ComparaisonHankel4}
\end{figure}

For concreteness, we switch now to explicit application of low order multipole solutions in order to get more physical insight into their properties. For the remainder of this paper, we focus on some illuminating cases such as the low order multipoles: dipole, quadrupole, hexapole and octopole.

\section{Approximate rotating multipole solutions}
\label{sec:ApproximateRotatingMultipole}

Including frame-dragging effects into the picture of rotating multipoles renders the problem analytically infeasible. The system of equations satisfied by the transformed potentials $u_{\ell,m}^{\rm D/B}$ has to be solved numerically. Confluent Heun equations and its generalization are solved by spectral methods, expanding the unknown functions onto Chebyshev polynomials. More precisely, elliptic equations satisfied by the potentials are efficiently solved by a radial expansion onto rational Chebyshev functions as defined and explained in \cite{boyd_chebyshev_2001}. Boundary conditions taking into account frame-dragging and space curvature have been given in eq.~(\ref{eq:LimitesfDlm}) and eq.~(\ref{eq:LimiteuBlm}). Moreover, these transformed potentials should converge to a finite non vanishing value at infinite radius.

In the following subsections, we given accurate numerical solutions to the electromagnetic field of a single rotating multipole up to octupole order~$\ell=4$. To emphasize the effect of space curvature and frame-dragging, we compute two kind of solutions. The first one assumes a background Schwarzschild metric thus without dragging and the second one includes this dragging adopting the slowly rotating approximation. Actually, we will show that except for unrealistically short periods with $P<1$~ms, this frame dragging can safely be ignored as far as the Poynting flux is concerned.

\cor{Because of frame-dragging effects that couple $f^{\rm B}_{\ell,m}$ modes to $f^{\rm D}_{\ell,m}$ modes according to the right-hand-side terms in eq.~(\ref{eq:Helmholtz}), there is an infinity of modes $(\ell,m)$ that are induced by a single rotating multipole. However, because these modes remain much weaker than the fundamental single multipole imposed by the star, we neglect them in the subsequent treatment. Time-dependent numerical simulations at the end of the present paper taking into account many multipoles $(\ell'>\ell,m')$ will confirm our expectations.}

\subsection{Dipole solution}

The rotating dipole has been extensively investigated in flat spacetime. If spacetime curvature is including, the transformed potentials have to satisfy the following generalization of the Helmholtz equation for a stationary wave. The two relevant functions are $u^{\rm D}_{2,1}$ and $u^{\rm B}_{1,1}$ verifying the coupled system
\begin{subequations}
\label{eq:HelmholtzDipole}
\begin{align}
 \frac{1}{r} \, \left[ r \, ( r - \Rs ) \, {u^{\rm D}_{2,1}}'' + ( 2 \, i \, k \, r^2 + \Rs ) \, {u^{\rm D}_{2,1}}' - 6 \, u^{\rm D}_{2,1} \right] + k^2 \, \left[ \left( 1 - \frac{\omega}{\Omega} \right)^2 - 1 \right] \, r \, \frac{u^{\rm D}_{2,1}}{\alpha^2} & = 3 \, \sqrt{\frac{3}{5}} \, \varepsilon_0 \, \omega \, u^{\rm B}_{1,1} \\
 \frac{1}{r} \, \left[ r \, ( r - \Rs ) \, {u^{\rm B}_{1,1}}'' + ( 2 \, i \, k \, r^2 + \Rs ) \, {u^{\rm B}_{1,1}}' - 2 \, u^{\rm B}_{1,1} \right] + k^2 \, \left[ \left( 1 - \frac{\omega}{\Omega} \right)^2 - 1 \right] \, r \, \frac{u^{\rm B}_{1,1}}{\alpha^2} & = 3 \, \sqrt{\frac{3}{5}} \, \mu_0 \, \omega \, u^{\rm D}_{2,1} .
\end{align}
\end{subequations}
Without frame-dragging that is when $\omega=0$, these equations decouple in \cor{two} scalar ordinary differential equations as is the case for Minkowski spacetime. In any case, the electric part has to satisfy the following boundary condition
\begin{equation}
 \alpha^2 \, \partial_r(u_{2,1}^{\rm D}\,e^{i\,k\,r_*}) = + \varepsilon_0 \, r \, \tilde{\omega} \, \sqrt{\frac{3}{5}} \, f_{1,1}^{\rm B} .
\end{equation}
The solution is completely and uniquely determined by the aforementioned equations and boundary conditions. In all subsequent results, as a check of the numerical implementation and high accuracy of our algorithm, we compare the output of the numerical approximation with the exact analytical solution expressed by spherical Hankel functions that are given by \cite{petri_multipolar_2015}.

For a dipole rotating in a Schwarzschild background metric, thus neglecting frame dragging, the spindown luminosity for different stellar compactness parameters~$\Xi=\Rs/R$ and different periods expressed by the ratio~$a=R/\rlight$ is shown in table~\ref{tab:DipoleSchwarzschild}. The column with $\Rs/R=0$ corresponds to the Minkowski solution and serves as a reference to prove the high precision of the method. Indeed, we found at least ten digits of precision compared to the analytical expression. This clearly points out the very good set of basis functions used to solve the system. As a general trend, the spindown monotonically increases with compactness for fixed period reaching up to 1.336 times the reference value.
\cor{As a comparison with the analytical expression found by \cite{rezzolla_electromagnetic_2004}, the corresponding general-relativistic enhancement is shown in the last line of the table. It depends on $\Rs/R$ but remains independent of the rotation rate so no need to specify $R/\rlight$. The formula is
\begin{equation}
\label{eq:Amplification}
\frac{L_{\rm GR}}{L_{\rm Newt}} = \left( \frac{f^{\rm B}_{1,0}(\Rs,R)}{f^{\rm B}_{1,0}(\Rs=0,R)} \right)^2 \, \left( 1 - \frac{\Rs}{R} \right)^{-2} 
\end{equation}
The first correction is induced by the magnetic field amplification and the second by gravitational redshift of the spin frequency. It should be a good guess for a point dipole, thus for $R/\rlight\ll1$. A quick comparison with the first line of the table with $R/\rlight=0.01$ clearly emphasizes the mismatch between both estimates. \cite{rezzolla_electromagnetic_2004} are severely overestimating the Poynting flux for very compact stars.
}
\begin{table}
 \centering
\begin{tabular}{ccccccc}
\hline
\diagbox{$R/\rlight$}{$\Rs/R$} & 0 & 0.1 & 0.2 & 0.3 & 0.4 & 0.5 \\
\hline
0.01 & 0.9999 & 1.003 & 1.006 & 1.009 & 1.012 & 1.015 \\
0.02 & 0.9996 & 1.005 & 1.012 & 1.018 & 1.025 & 1.032 \\
0.05 & 0.9975 & 1.013 & 1.029 & 1.045 & 1.063 & 1.081 \\
0.1  & 0.9901 & 1.020 & 1.053 & 1.088 & 1.125 & 1.166 \\
0.2  & 0.9615 & 1.018 & 1.083 & 1.156 & 1.239 & 1.336 \\
\cor{R.A.} & 1.000 & 1.444 & 2.171 & 3.434 & 5.800 & 10.70 \\
\hline
\end{tabular} 
\caption{Normalized spindown luminosity for the perpendicular rotating dipole in Schwarzschild background. The analytical approximation eq.~(\ref{eq:Amplification}) is shown in the last line (R.A.).}
\label{tab:DipoleSchwarzschild}
\end{table} 

\begin{figure}
\centering
\input{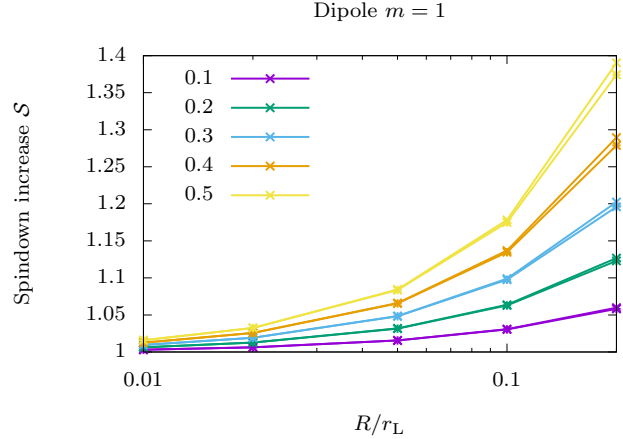}
\caption{Spindown luminosity of the orthogonal dipole in Schwarzschild and slowly rotating neutron star metric approximation.}
\label{fig:Dipole}
\end{figure}

When frame dragging is switch on, the Poynting flux decreases slightly compared to the previous case as seen in table~\ref{tab:DipoleSRNS}. This trend is easily explained by the boundary conditions for the electric field which contain the term $\tilde{\omega}$ that is a decrease in the effective rotation rate of the star as measured by a local observer explaining a weaker radiating electric field contribution. However, for realistic neutron star parameters, we would expect at most $R/\rlight\leqslant0.1$ and $\Rs/R\leqslant0.4$. In that parameter range, the discrepancy between table~\ref{tab:DipoleSchwarzschild} and table~\ref{tab:DipoleSRNS} remains less than 1\%. Thus for practical purposes when discussing the spindown luminosities, frame dragging effects are irrelevant.
\begin{table}
 \centering
\begin{tabular}{ccccccc}
\hline
\diagbox{$R/\rlight$}{$\Rs/R$} & 0 & 0.1 & 0.2 & 0.3 & 0.4 & 0.5 \\
\hline
0.01 & 0.9999 & 1.003 & 1.006 & 1.009 & 1.012 & 1.015 \\
0.02 & 0.9996 & 1.005 & 1.012 & 1.018 & 1.025 & 1.031 \\
0.05 & 0.9975 & 1.012 & 1.028 & 1.045 & 1.062 & 1.080 \\
0.1 & 0.9901 & 1.020 & 1.052 & 1.086 & 1.123 & 1.163 \\
0.2 & 0.9615 & 1.017 & 1.079 & 1.149 & 1.229 & 1.321 \\
\hline
\end{tabular} 
\caption{Normalized spindown luminosity for the perpendicular rotating dipole in the slowly rotating metric background.}
\label{tab:DipoleSRNS}
\end{table} 
The plots in fig.\ref{fig:Dipole} give a synthetic overview of the full set of results showing the increase in spindown luminosity compared to flat spacetime given by the ratio
\begin{equation}
 \mathcal{S}(a,\Xi) = \frac{L_{\rm GR}}{L_{\rm flat}} .
\end{equation}
Each curve corresponds to the ratio between a compactness~$\Xi>0$ and the reference case $\Xi=0$. There is always an increase in the Poynting flux whatever the period~$a$ and compactness~$\Xi$. \cor{Solid lines with the same color correspond to a same compactness. The Schwarzschild luminosity always lies above the slowly rotating neutron star metric. This is also valid for the higher multipoles shown in the following paragraphs.}

\cor{To emphasize the difference between the numerical solution and some analytical approximate expressions, we plot the real and imaginary part of the radial dependence of the magnetic field for a dipole in Schwarzschild spacetime with the parameters $R/\rlight=0.1$ and $R/\Rs=2$. The discrepancies are highlighted in fig.~\ref{fig:Hankel_Reel}. The numerical solution is shown in violet solid line (num in the legend). The expression deduced according to \cite{rezzolla_electromagnetic_2004} does not reproduce the solution to good accuracy, green solid line (R.A. in the legend). Actually, we found that the Newtonian expression, in solid blue line ($h_1^{(1)}$ in the legend), assuming variations according to spherical Hankel functions $h_\ell^{(1)}(k_m\,r)$, regardless of any general-relativistic effect such as gravitation redshift or stellar boundary condition, gives the best approximation to the numerical solution. Corrections are only significant for unrealistically high rotation rates $R/\rlight\lesssim1$. Our conclusions remain valid for a slowly rotating spacetime metric as long as $R/\rlight\ll1$. This explains why the general-relativistic luminosity equals approximately the Minkowski results. For a distant observer, there is no distinction between general-relativistic or Minkowski metric. Obviously, for an observer sitting on the star, the measurements would be very different.
\begin{figure}
\centering
\input{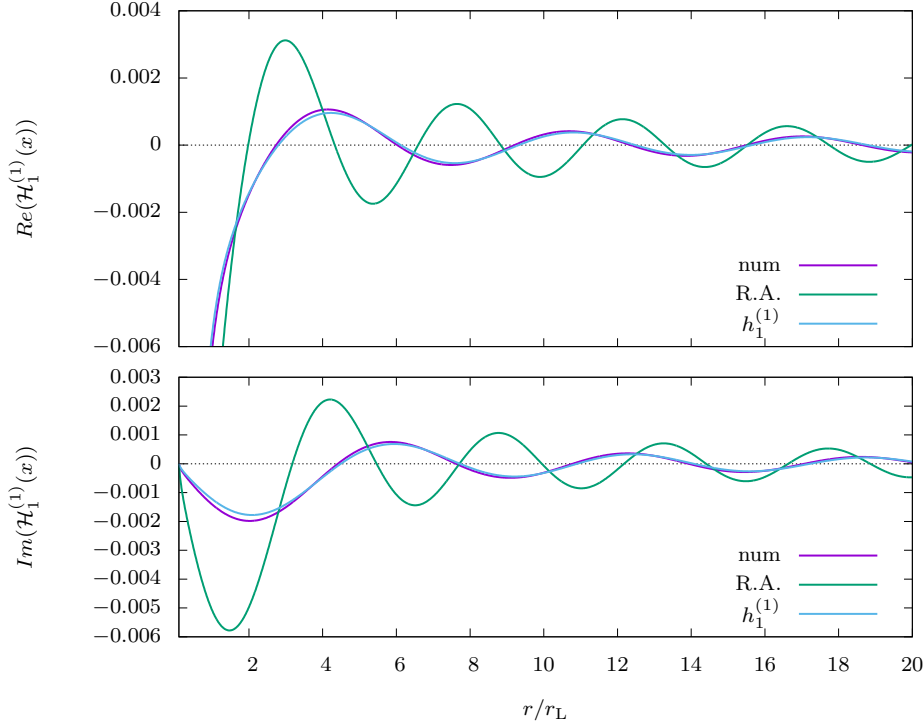}
\caption{Comparison between the numerical, red solid line, and the analytical expressions, green and blue solid lines, for the radial dependence of the magnetic field profile. The parameters are $R/\rlight=0.1$ and $R/\Rs=2$ for Schwarzschild spacetime.}
\label{fig:Hankel_Reel}
\end{figure}
}

\subsection{Quadrupole solution}

The rotating quadrupole has been less extensively investigated, even in flat spacetime. The three relevant transformed potentials for the $m=1$ mode are $u^{\rm D}_{1,1}, u^{\rm D}_{3,1}$ and $u^{\rm B}_{2,1}$. They verify the coupled system for the three unknown functions 
\begin{subequations}
\label{eq:HelmholtzQuadrupoleM1}
\begin{align}
 \frac{1}{r} \, \left[ r \, ( r - \Rs ) \, {u^{\rm D}_{1,1}}'' + ( 2 \, i \, k \, r^2 + \Rs ) \, {u^{\rm D}_{1,1}}' - 2 \, u^{\rm D}_{1,1} \right] + k^2 \, \left[ \left( 1 - \frac{\omega}{\Omega} \right)^2 - 1 \right] \, r \, \frac{u^{\rm D}_{1,1}}{\alpha^2} & = - 3 \, \sqrt{\frac{3}{5}} \, \varepsilon_0 \, \omega \, u^{\rm B}_{2,1} \\
 \frac{1}{r} \, \left[ r \, ( r - \Rs ) \, {u^{\rm B}_{2,1}}'' + ( 2 \, i \, k \, r^2 + \Rs ) \, {u^{\rm B}_{2,1}}' - 6 \, u^{\rm B}_{2,1} \right] + k^2 \, \left[ \left( 1 - \frac{\omega}{\Omega} \right)^2 - 1 \right] \, r \, \frac{u^{\rm B}_{2,1}}{\alpha^2} & =\\
 - 3 \, \mu_0 \, \omega \, \left( \sqrt{\frac{3}{5}} \, u^{\rm D}_{1,1} - \frac{8}{\sqrt{35}} \, u^{\rm D}_{3,1} \right) & \\
 \frac{1}{r} \, \left[ r \, ( r - \Rs ) \, {u^{\rm D}_{3,1}}'' + ( 2 \, i \, k \, r^2 + \Rs ) \, {u^{\rm D}_{3,1}}' - 12 \, u^{\rm D}_{3,1} \right] + k^2 \, \left[ \left( 1 - \frac{\omega}{\Omega} \right)^2 - 1 \right] \, r \, \frac{u^{\rm D}_{3,1}}{\alpha^2} & = 3 \, \frac{8}{\sqrt{35}} \, \varepsilon_0 \, \omega \, u^{\rm B}_{2,1} \ .
\end{align}
\end{subequations}
The appropriate boundary conditions for the electric part are given by
\begin{subequations}
\begin{align}
 \alpha^2 \, \partial_r(u_{1,1}^{\rm D}\,e^{i\,k\,r_*}) & = - \varepsilon_0 \, r \, \tilde{\omega} \, \sqrt{\frac{3}{5}} \, f_{2,1}^{\rm B} \\
 \alpha^2 \, \partial_r(u_{3,1}^{\rm D}\,e^{i\,k\,r_*}) & = + \varepsilon_0 \, r \, \tilde{\omega} \, \frac{8}{\sqrt{35}} \, f_{2,1}^{\rm B} .
\end{align}
\end{subequations}
The spindown luminosity obtained from the numerical integration of this coupled system is summarized in table~\ref{tab:QuadrupoleM1Schwarzschild} for the Schwarzschild metric. Here again, as a check the flat spacetime results are compared to analytical expressions and the precision remains better than 10~digits. The luminosity increase with respect to compactness is more sensitive than for the dipole $m=1$ mode up to 6 times larger than the flat spacetime quadrupole.
\begin{table}
 \centering
\begin{tabular}{ccccccc}
\hline
\diagbox{$R/\rlight$}{$\Rs/R$} & 0 & 0.1 & 0.2 & 0.3 & 0.4 & 0.5 \\
\hline
0.01 & 1.000 & 1.164 & 1.382 & 1.680 & 2.105 & 2.742 \\
0.02 & 1.000 & 1.168 & 1.391 & 1.697 & 2.132 & 2.788 \\
0.05 & 1.001 & 1.181 & 1.420 & 1.749 & 2.221 & 2.934 \\
0.1  & 1.007 & 1.206 & 1.474 & 1.846 & 2.386 & 3.214 \\
0.2  & 1.031 & 1.272 & 1.604 & 2.081 & 2.795 & 3.929 \\
\hline
\end{tabular} 
\caption{Normalized spindown luminosity for the $m=1$ rotating quadrupole in Schwarzschild background.}
\label{tab:QuadrupoleM1Schwarzschild}
\end{table} 
If frame-dragging is switch on, the decrease in luminosity can be quantified by comparing table~\ref{tab:QuadrupoleM1Schwarzschild} with table~\ref{tab:QuadrupoleM1SRNS}. We always observe a lowering of the spindown rate because of the boundary conditions imposed on the electric part for the same reason as for the dipole. Frame dragging decrease the effective stellar rotation rate for a local observer and thus the expected electric field strength contribution to the radiating field.
\begin{table}
 \centering
\begin{tabular}{ccccccc}
\hline
\diagbox{$R/\rlight$}{$\Rs/R$} & 0 & 0.1 & 0.2 & 0.3 & 0.4 & 0.5 \\
\hline
0.01 & 1.000 & 1.136 & 1.312 & 1.545 & 1.864 & 2.323 \\
0.02 & 1.000 & 1.140 & 1.321 & 1.560 & 1.888 & 2.362 \\
0.05 & 1.001 & 1.152 & 1.348 & 1.607 & 1.965 & 2.483 \\
0.1  & 1.007 & 1.176 & 1.397 & 1.693 & 2.105 & 2.708 \\
0.2  & 1.031 & 1.238 & 1.513 & 1.892 & 2.434 & 3.250 \\
\hline
\end{tabular} 
\caption{Normalized spindown luminosity for the $m=1$ rotating quadrupole in the slowly rotating background metric.}
\label{tab:QuadrupoleM1SRNS}
\end{table}
The plots in fig.\ref{fig:QuadrupoleM1} give a synthetic compilation of the results showing the increase in spindown luminosity compared to flat spacetime reference values.

\begin{figure}
\centering
\input{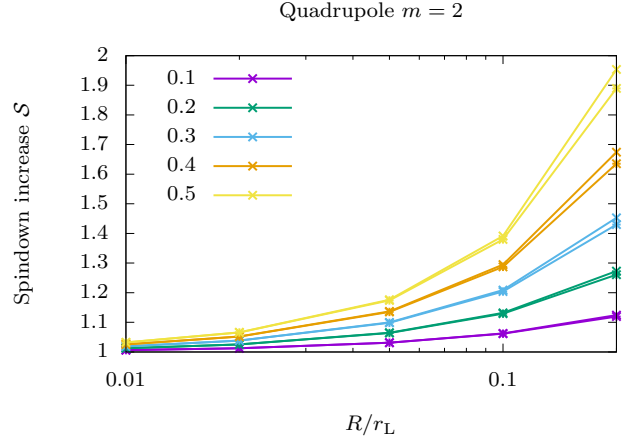}
\caption{Spindown luminosity of the orthogonal rotator quadrupole for the mode $m=2$ in Schwarzschild and slowly rotating neutron star metric approximation.}
\label{fig:QuadrupoleM2}
\end{figure}

For the $m=2$ the two relevant transformed potentials are $u^{\rm D}_{3,2}$ and $u^{\rm B}_{2,2}$. They verify the coupled system of two unknown functions
\begin{subequations}
\label{eq:HelmholtzQuadrupoleM2}
\begin{align}
 \frac{1}{r} \, \left[ r \, ( r - \Rs ) \, {u^{\rm B}_{2,2}}'' + ( 2 \, i \, k_2 \, r^2 + \Rs ) \, {u^{\rm B}_{2,2}}' - 6 \, u^{\rm B}_{2,2} \right] + k_2^2 \, \left[ \left( 1 - \frac{\omega}{\Omega} \right)^2 - 1 \right] \, r \, \frac{u^{\rm B}_{2,2}}{\alpha^2} & = 6 \, \sqrt{\frac{2}{7}} \, \mu_0 \, \omega \, u^{\rm D}_{3,2} \\
 \frac{1}{r} \, \left[ r \, ( r - \Rs ) \, {u^{\rm D}_{3,2}}'' + ( 2 \, i \, k_2 \, r^2 + \Rs ) \, {u^{\rm D}_{3,2}}' - 12 \, u^{\rm D}_{3,2} \right] + k_2^2 \, \left[ \left( 1 - \frac{\omega}{\Omega} \right)^2 - 1 \right] \, r \, \frac{u^{\rm D}_{3,2}}{\alpha^2} & = 6 \, \frac{\sqrt{14}}{7} \, \varepsilon_0 \, \omega \, u^{\rm B}_{2,2} \ .
\end{align}
\end{subequations}
The boundary conditions apply on the electric field such that
\begin{subequations}
\begin{align}
 \alpha^2 \, \partial_r(u_{3,2}^{\rm D}\,e^{i\,k_2\,r_*}) & = 2 \, \sqrt{\frac{2}{7}} \, \varepsilon_0 \, r \, \tilde{\omega} \, f_{2,1}^{\rm B} .
\end{align}
\end{subequations}
Table~\ref{tab:QuadrupoleM2Schwarzschild} summarizes results for the Schwarzschild metric and table~\ref{tab:QuadrupoleM2SRNS} allows comparison with the slowly rotating neutron star metric. Compactness increases again the spindown rate although it is less pronounced than for the $m=1$ mode.
\begin{table}
 \centering
\begin{tabular}{ccccccc}
\hline
\diagbox{$R/\rlight$}{$\Rs/R$} & 0 & 0.1 & 0.2 & 0.3 & 0.4 & 0.5 \\
\hline
0.01 & 0.9998 & 1.006  & 1.012 & 1.0182 & 1.025 & 1.031 \\
0.02 & 0.9994 & 1.012  & 1.024 & 1.0387 & 1.051 & 1.065 \\
0.05 & 0.9966 & 1.028  & 1.061 & 1.0967 & 1.133 & 1.172 \\
0.1  & 0.9866 & 1.048  & 1.117 & 1.1924 & 1.277 & 1.372 \\
0.2  & 0.9468 & 1.064  & 1.205 & 1.3755 & 1.585 & 1.849 \\
\hline
\end{tabular} 
\caption{Normalized spindown luminosity for the $m=2$ rotating quadrupole in Schwarzschild background.}
\label{tab:QuadrupoleM2Schwarzschild}
\end{table} 
\begin{table}
 \centering
\begin{tabular}{ccccccc}
\hline
\diagbox{$R/\rlight$}{$\Rs/R$} & 0 & 0.1 & 0.2 & 0.3 & 0.4 & 0.5 \\
\hline
0.01 & 0.9998 & 1.006 & 1.012 & 1.018 & 1.025 & 1.031 \\
0.02 & 0.9994 & 1.012 & 1.024 & 1.037 & 1.051 & 1.065 \\
0.05 & 0.9966 & 1.027 & 1.060 & 1.094 & 1.131 & 1.170 \\
0.1  & 0.9866 & 1.047 & 1.114 & 1.188 & 1.269 & 1.361 \\
0.2  & 0.9468 & 1.059 & 1.193 & 1.353 & 1.548 & 1.788 \\
\hline
\end{tabular} 
\caption{Normalized spindown luminosity for the $m=2$ rotating quadrupole in the slowly rotating background metric.}
\label{tab:QuadrupoleM2SRNS}
\end{table} 
The plots in fig.\ref{fig:QuadrupoleM2} give a synthetic compilation of the results showing the increase in spindown luminosity compared to flat spacetime.

\begin{figure}
\centering
\input{spindown_quadrupole_m1_NR30.tex}
\caption{Spindown luminosity of the orthogonal quadrupole for the mode $m=1$ in Schwarzschild and slowly rotating neutron star metric approximation.}
\label{fig:QuadrupoleM1}
\end{figure}

\subsection{Hexapole solution}

Higher order multipoles $\ell\geqslant2$ are treated in the same way as a dipole or quadrupole. There is nor increase in the complexity of the algorithm or in the numerical computation of the Poynting flux. However, the possible parameter space for the geometric configuration of the magnetic field augments with $\ell$ because there is an increasing number of modes~$m$ such that $m\leqslant\ell$. High order multipoles are useful to investigate small scale structure on the neutron star surface and also to look for off-centred dipole solutions \citep{petri_radiation_2016}. In order to quantitatively fix the modification brought by general relativity, we decide two give results up to the octupole. 

The required equations for the $m=1$ hexapole mode involving $u^{D}_{2,1}, u^{D}_{4,1}, u^{\rm B}_{3,1}$ are
\begin{subequations}
\label{eq:HelmholtzHexapoleM1}
\begin{align}
 \frac{1}{r} \, \left[ r \, ( r - \Rs ) \, {u^{D}_{2,1}}'' + ( 2 \, i \, k \, r^2 + \Rs ) \, {u^{D}_{2,1}}' - 6 \, u^{D}_{2,1} \right] + k^2 \, \left[ \left( 1 - \frac{\omega}{\Omega} \right)^2 - 1 \right] \, r \, \frac{u^{D}_{2,1}}{\alpha^2} & = - 3 \, \frac{8}{\sqrt{35}} \, \varepsilon_0 \, \omega \, u^{\rm B}_{3,1} \\
 \frac{1}{r} \, \left[ r \, ( r - \Rs ) \, {u^{\rm B}_{3,1}}'' + ( 2 \, i \, k \, r^2 + \Rs ) \, {u^{\rm B}_{3,1}}' - 12 \, u^{\rm B}_{3,1} \right] + k^2 \, \left[ \left( 1 - \frac{\omega}{\Omega} \right)^2 - 1 \right] \, r \, \frac{u^{\rm B}_{3,1}}{\alpha^2} & \\
 = - 3 \, \mu_0 \, \omega \, \left( \frac{8}{\sqrt{35}} \, u^{\rm D}_{2,1} - \frac{5}{\sqrt{7}} \, u^{\rm D}_{4,1} \right) & \\
 \frac{1}{r} \, \left[ r \, ( r - \Rs ) \, {u^{D}_{4,1}}'' + ( 2 \, i \, k \, r^2 + \Rs ) \, {u^{D}_{4,1}}' - 20 \, u^{D}_{4,1} \right] + k^2 \, \left[ \left( 1 - \frac{\omega}{\Omega} \right)^2 - 1 \right] \, r \, \frac{u^{D}_{4,1}}{\alpha^2} & = 3 \, \frac{5}{\sqrt{7}} \, \varepsilon_0 \, \omega \, u^{\rm B}_{3,1} \ .
\end{align}
\end{subequations}
The boundary conditions for the electric field are
\begin{subequations}
\begin{align}
 \alpha^2 \, \partial_r(u_{2,1}^{\rm D}\,e^{i\,k\,r_*}) & = - \frac{8}{\sqrt{35}} \, \varepsilon_0 \, r \, \tilde{\omega} \, f_{3,1}^{\rm B} \\
 \alpha^2 \, \partial_r(u_{4,1}^{\rm D}\,e^{i\,k\,r_*}) & = + \frac{5}{\sqrt{7}} \, \varepsilon_0 \, r \, \tilde{\omega} \, f_{3,1}^{\rm B} .
\end{align}
\end{subequations}
Applying the same procedure as before, table~\ref{tab:HexapoleM1Schwarzschild} show the luminosity we obtained for the Schwarzschild metric. The increase in luminosity reaches values up to 7.
\begin{table}
 \centering
\begin{tabular}{ccccccc}
\hline
\diagbox{$R/\rlight$}{$\Rs/R$} & 0 & 0.1 & 0.2 & 0.3 & 0.4 & 0.5 \\
\hline
0.01 & 0.9999 & 1.188 & 1.441 & 1.791 & 2.296 & 3.066 \\
0.02 & 0.9999 & 1.192 & 1.450 & 1.807 & 2.324 & 3.113 \\
0.05 & 0.9999 & 1.203 & 1.477 & 1.860 & 2.415 & 3.268 \\
0.1  & 0.9998 & 1.222 & 1.525 & 1.951 & 2.578 & 3.552 \\
0.2  & 0.9994 & 1.259 & 1.622 & 2.148 & 2.943 & 4.216 \\
\hline
\end{tabular} 
\caption{Normalized spindown luminosity for the $m=1$ rotating hexapole in Schwarzschild background.}
\label{tab:HexapoleM1Schwarzschild}
\end{table}
Comparison with the slowly rotating neutron star metric is always useful and given in  table~\ref{tab:HexapoleM1SRNS}.
\begin{table}
 \centering
\begin{tabular}{ccccccc}
\hline
\diagbox{$R/\rlight$}{$\Rs/R$} & 0 & 0.1 & 0.2 & 0.3 & 0.4 & 0.5 \\
\hline
0.01 & 0.9999 & 1.154 & 1.354 & 1.622 & 1.994 & 2.536 \\
0.02 & 0.9999 & 1.158 & 1.363 & 1.637 & 2.018 & 2.574 \\
0.05 & 0.9999 & 1.168 & 1.388 & 1.684 & 2.096 & 2.701 \\
0.1  & 0.9998 & 1.186 & 1.432 & 1.765 & 2.234 & 2.929 \\
0.2  & 0.9994 & 1.221 & 1.520 & 1.935 & 2.534 & 3.445 \\
\hline
\end{tabular} 
\caption{Normalized spindown luminosity for the $m=1$ rotating hexapole in a slowly rotating metric background.}
\label{tab:HexapoleM1SRNS}
\end{table} 
The plots in fig.\ref{fig:HexapoleM1} give a synthetic compilation of the results showing the increase in spindown luminosity compared to flat spacetime.

If we consider the $m=2$ hexapole mode, the appropriate functions are $u^{D}_{2,2}, u^{D}_{4,2}, u^{\rm B}_{3,2}$ that satisfy
\begin{subequations}
\label{eq:HelmholtzHexapoleM2}
\begin{align}
 \frac{1}{r} \, \left[ r \, ( r - \Rs ) \, {u^{D}_{2,2}}'' + ( 2 \, i \, k_2 \, r^2 + \Rs ) \, {u^{D}_{2,2}}' - 6 \, u^{D}_{2,2} \right] + k_2^2 \, \left[ \left( 1 - \frac{\omega}{\Omega} \right)^2 - 1 \right] \, r \, \frac{u^{D}_{2,2}}{\alpha^2} & = -6 \, \sqrt{\frac{2}{7}} \, \varepsilon_0 \, \omega \, u^{\rm B}_{3,2} \\
 \frac{1}{r} \, \left[ r \, ( r - \Rs ) \, {u^{\rm B}_{3,2}}'' + ( 2 \, i \, k_2 \, r^2 + \Rs ) \, {u^{\rm B}_{3,2}}' - 12 \, u^{\rm B}_{3,2} \right] + k_2^2 \, \left[ \left( 1 - \frac{\omega}{\Omega} \right)^2 - 1 \right] \, r \, \frac{u^{\rm B}_{3,2}}{\alpha^2} & = \\
 - 6 \, \mu_0 \, \omega \, \left( \frac{\sqrt{14}}{7} \, u^{\rm D}_{2,2} - \frac{\sqrt{35}}{7} \, u^{\rm D}_{4,2} \right) \\
 \frac{1}{r} \, \left[ r \, ( r - \Rs ) \, {u^{D}_{4,2}}'' + ( 2 \, i \, k_2 \, r^2 + \Rs ) \, {u^{D}_{4,2}}' - 20 \, u^{D}_{4,2} \right] + k_2^2 \, \left[ \left( 1 - \frac{\omega}{\Omega} \right)^2 - 1 \right] \, r \, \frac{u^{D}_{4,2}}{\alpha^2} & = 6 \, \sqrt{\frac{5}{7}} \, \varepsilon_0 \, \omega \, u^{\rm B}_{3,2} \ .
\end{align}
\end{subequations}
The boundary conditions for the electric field are
\begin{subequations}
\begin{align}
 \alpha^2 \, \partial_r(u_{2,2}^{\rm D}\,e^{i\,k_2\,r_*}) & = - 2 \, \sqrt{\frac{2}{7}} \, \varepsilon_0 \, r \, \tilde{\omega} \, f_{3,2}^{\rm B} \\
 \alpha^2 \, \partial_r(u_{4,2}^{\rm D}\,e^{i\,k_2\,r_*}) & = 2 \, \sqrt{\frac{5}{7}} \, \varepsilon_0 \, r \, \tilde{\omega} \, f_{3,2}^{\rm B} .
\end{align}
\end{subequations}
The accurate value of the spindown for Schwarzschild metric are given in  table~\ref{tab:HexapoleM2Schwarzschild} and for the slowly rotating neutron star in table~\ref{tab:HexapoleM2SRNS}.
\begin{table}
 \centering
\begin{tabular}{ccccccc}
\hline
\diagbox{$R/\rlight$}{$\Rs/R$} & 0 & 0.1 & 0.2 & 0.3 & 0.4 & 0.5 \\
\hline
0.01 & 0.9999 & 1.136 & 1.317 & 1.568 & 1.931 & 2.483 \\
0.02 & 0.9998 & 1.143 & 1.334 & 1.598 & 1.980 & 2.564 \\
0.05 & 0.9992 & 1.163 & 1.385 & 1.692 & 2.140 & 2.831 \\
0.1  & 0.9972 & 1.197 & 1.471 & 1.861 & 2.440 & 3.356 \\
0.2  & 0.9899 & 1.261 & 1.653 & 2.242 & 3.178 & 4.771 \\
\hline
\end{tabular} 
\caption{Normalized spindown luminosity for the $m=2$ rotating hexapole in Schwarzschild background.}
\label{tab:HexapoleM2Schwarzschild}
\end{table}
\begin{table}
 \centering
\begin{tabular}{ccccccc}
\hline
\diagbox{$R/\rlight$}{$\Rs/R$} & 0 & 0.1 & 0.2 & 0.3 & 0.4 & 0.5 \\
\hline
0.01 & 0.9999 & 1.112 & 1.257 & 1.450 & 1.717 & 2.107 \\
0.02 & 0.9998 & 1.118 & 1.272 & 1.477 & 1.761 & 2.176 \\
0.05 & 0.9992 & 1.139 & 1.320 & 1.562 & 1.900 & 2.396 \\
0.1  & 0.9972 & 1.170 & 1.399 & 1.711 & 2.154 & 2.816 \\
0.2  & 0.9899 & 1.228 & 1.557 & 2.029 & 2.737 & 3.863 \\
\hline
\end{tabular} 
\caption{Normalized spindown luminosity for the $m=2$ rotating hexapole in a slowly rotating metric background.}
\label{tab:HexapoleM2SRNS}
\end{table} 
The plots in fig.\ref{fig:HexapoleM2} give a synthetic compilation of the results showing the increase in spindown luminosity compared to flat spacetime.

Finally, the $m=3$ hexapole mode requires the solution of $u^{D}_{4,3}, u^{\rm B}_{3,3}$ satisfying
\begin{subequations}
\label{eq:HelmholtzHexapoleM3}
\begin{align}
 \frac{1}{r} \, \left[ r \, ( r - \Rs ) \, {u^{\rm B}_{3,3}}'' + ( 2 \, i \, k_3 \, r^2 + \Rs ) \, {u^{\rm B}_{3,3}}' - 12 \, u^{\rm B}_{3,3} \right] + k_3^2 \, \left[ \left( 1 - \frac{\omega}{\Omega} \right)^2 - 1 \right] \, r \, \frac{u^{\rm B}_{3,3}}{\alpha^2} & = \sqrt{15} \, \mu_0 \, \omega \, u^{\rm D}_{4,3} \\
 \frac{1}{r} \, \left[ r \, ( r - \Rs ) \, {u^{D}_{4,3}}'' + ( 2 \, i \, k_3 \, r^2 + \Rs ) \, {u^{D}_{4,3}}' - 20 \, u^{D}_{4,3} \right] + k_3^2 \, \left[ \left( 1 - \frac{\omega}{\Omega} \right)^2 - 1 \right] \, r \, \frac{u^{D}_{4,3}}{\alpha^2} & = 3 \, \sqrt{\frac{5}{3}} \, \varepsilon_0 \, \omega \, u^{\rm B}_{3,3}
\end{align}
\end{subequations}
supplemented with the boundary conditions for the electric field as
\begin{subequations}
\begin{align}
 \alpha^2 \, \partial_r(u_{4,3}^{\rm D}\,e^{i\,k_3\,r_*}) & = \sqrt{\frac{5}{3}} \, \varepsilon_0 \, r \, \tilde{\omega} \, f_{3,3}^{\rm B} .
\end{align}
\end{subequations}
Comparison of table~\ref{tab:HexapoleM3Schwarzschild} with table~\ref{tab:HexapoleM3SRNS} demonstrates again the decrease of spindown induced by the frame dragging effect.
\begin{table}
 \centering
\begin{tabular}{ccccccc}
\hline
\diagbox{$R/\rlight$}{$\Rs/R$} & 0 & 0.1 & 0.2 & 0.3 & 0.4 & 0.5 \\
\hline
0.01 & 0.9998 & 1.009 & 1.018 & 1.028 & 1.038 & 1.048 \\
0.02 & 0.9992 & 1.018 & 1.037 & 1.057 & 1.078 & 1.100 \\
0.05 & 0.9955 & 1.042 & 1.093 & 1.148 & 1.207 & 1.270 \\
0.1  & 0.9821 & 1.076 & 1.183 & 1.305 & 1.447 & 1.614 \\
0.2  & 0.9296 & 1.108 & 1.335 & 1.630 & 2.020 & 2.551 \\
\hline
\end{tabular} 
\caption{Normalized spindown luminosity for the $m=3$ rotating hexapole in Schwarzschild background.}
\label{tab:HexapoleM3Schwarzschild}
\end{table}
\begin{table}
 \centering
\begin{tabular}{ccccccc}
\hline
\diagbox{$R/\rlight$}{$\Rs/R$} & 0 & 0.1 & 0.2 & 0.3 & 0.4 & 0.5 \\
\hline
0.01 & 0.9998 & 1.009 & 1.018 & 1.028 & 1.038 & 1.048 \\
0.02 & 0.9992 & 1.018 & 1.037 & 1.057 & 1.078 & 1.099 \\
0.05 & 0.9955 & 1.042 & 1.092 & 1.146 & 1.204 & 1.266 \\
0.1  & 0.9821 & 1.074 & 1.178 & 1.297 & 1.433 & 1.591 \\
0.2  & 0.9296 & 1.100 & 1.313 & 1.586 & 1.940 & 2.409 \\
\hline
\end{tabular} 
\caption{Normalized spindown luminosity for the $m=3$ rotating quadrupole in a slowly rotating metric background.}
\label{tab:HexapoleM3SRNS}
\end{table} 
The plots in fig.\ref{fig:HexapoleM3} give a synthetic compilation of the results showing the increase in spindown luminosity compared to flat spacetime.

\begin{figure}
\centering
\input{spindown_hexapole_m1_NR30.tex}
\caption{Spindown luminosity of the orthogonal rotator hexapole for the mode $m=1$ in Schwarzschild and slowly rotating neutron star metric approximation.}
\label{fig:HexapoleM1}
\end{figure}

\begin{figure}
\centering
\input{spindown_hexapole_m2_NR30.tex}
\caption{Spindown luminosity of the orthogonal rotator hexapole for the mode $m=2$ in Schwarzschild and slowly rotating neutron star metric approximation.}
\label{fig:HexapoleM2}
\end{figure}

\begin{figure}
\centering
\input{spindown_hexapole_m3_NR30.tex}
\caption{Spindown luminosity of the orthogonal rotator hexapole for the mode $m=3$ in Schwarzschild and slowly rotating neutron star metric approximation.}
\label{fig:HexapoleM3}
\end{figure}

\subsection{Octupole solution}

Eventually, the octupole can be solved following the same lines. For the $m=1$ octupole mode, the coupled system reads for $u^{D}_{3,1}, u^{D}_{5,1}, u^{\rm B}_{4,1}$
\begin{subequations}
\label{eq:HelmholtzOctupoleM1}
\begin{align}
 \frac{1}{r} \, \left[ r \, ( r - \Rs ) \, {u^{D}_{3,1}}'' + ( 2 \, i \, k \, r^2 + \Rs ) \, {u^{D}_{3,1}}' - 12 \, u^{D}_{3,1} \right] + k^2 \, \left[ \left( 1 - \frac{\omega}{\Omega} \right)^2 - 1 \right] \, r \, \frac{u^{D}_{3,1}}{\alpha^2} & = - 3 \, \frac{5}{\sqrt{7}} \, \varepsilon_0 \, \omega \, u^{\rm B}_{4,1} \\
 \frac{1}{r} \, \left[ r \, ( r - \Rs ) \, {u^{\rm B}_{4,1}}'' + ( 2 \, i \, k \, r^2 + \Rs ) \, {u^{\rm B}_{4,1}}' - 20 \, u^{\rm B}_{4,1} \right] + k^2 \, \left[ \left( 1 - \frac{\omega}{\Omega} \right)^2 - 1 \right] \, r \, \frac{u^{\rm B}_{4,1}}{\alpha^2} & = \\
 - 3 \, \mu_0 \, \omega \, \left( \frac{5}{\sqrt{7}} \, f^{\rm D}_{3,1} - \frac{8}{\sqrt{11}} \, f^{\rm D}_{5,1} \right) \\
 \frac{1}{r} \, \left[ r \, ( r - \Rs ) \, {u^{D}_{5,1}}'' + ( 2 \, i \, k \, r^2 + \Rs ) \, {u^{D}_{5,1}}' - 30 \, u^{D}_{5,1} \right] + k^2 \, \left[ \left( 1 - \frac{\omega}{\Omega} \right)^2 - 1 \right] \, r \, \frac{u^{D}_{5,1}}{\alpha^2} & = 3 \, \frac{8}{\sqrt{11}} \, \varepsilon_0 \, \omega \, u^{\rm B}_{4,1}
\end{align}
\end{subequations}
with the boundary conditions for the electric field
\begin{subequations}
\begin{align}
 \alpha^2 \, \partial_r(u_{3,1}^{\rm D}\,e^{i\,k\,r_*}) & = - \frac{5}{\sqrt{7}} \, \varepsilon_0 \, r \, \tilde{\omega} \, f_{4,1}^{\rm B} \\
 \alpha^2 \, \partial_r(u_{5,1}^{\rm D}\,e^{i\,k\,r_*}) & = + \frac{8}{\sqrt{11}} \, \varepsilon_0 \, r \, \tilde{\omega} \, f_{4,1}^{\rm B} .
\end{align}
\end{subequations}
Performing the numerical integration results are shown in table~\ref{tab:OctupoleM1Schwarzschild} for Schwarzschild metric and in table~\ref{tab:OctupoleM1SRNS} for the slowly rotating neutron star metric.
\begin{table}
 \centering
\begin{tabular}{ccccccc}
\hline
\diagbox{$R/\rlight$}{$\Rs/R$} & 0 & 0.1 & 0.2 & 0.3 & 0.4 & 0.5 \\
\hline
0.01 & 0.9999 & 1.199 & 1.468 & 1.844 & 2.390 & 3.231 \\
0.02 & 0.9999 & 1.204 & 1.481 & 1.867 & 2.429 & 3.295 \\
0.05 & 0.9998 & 1.215 & 1.509 & 1.920 & 2.522 & 3.455 \\
0.1  & 0.9992 & 1.234 & 1.556 & 2.013 & 2.689 & 3.748 \\
0.2  & 0.9971 & 1.270 & 1.653 & 2.212 & 3.060 & 4.427 \\
\hline
\end{tabular} 
\caption{Normalized spindown luminosity for the $m=1$ rotating octupole in Schwarzschild background.}
\label{tab:OctupoleM1Schwarzschild}
\end{table} 
\begin{table}
 \centering
\begin{tabular}{ccccccc}
\hline
\diagbox{$R/\rlight$}{$\Rs/R$} & 0 & 0.1 & 0.2 & 0.3 & 0.4 & 0.5 \\
\hline
0.01 & 0.9999 & 1.161 & 1.373 & 1.656 & 2.053 & 2.636 \\
0.02 & 0.9999 & 1.166 & 1.384 & 1.677 & 2.086 & 2.689 \\
0.05 & 0.9998 & 1.177 & 1.410 & 1.724 & 2.166 & 2.818 \\
0.1  & 0.9992 & 1.195 & 1.454 & 1.807 & 2.307 & 3.053 \\
0.2  & 0.9971 & 1.229 & 1.542 & 1.980 & 2.614 & 3.585 \\
\hline
\end{tabular} 
\caption{Normalized spindown luminosity for the $m=1$ rotating octupole in a slowly rotating metric background.}
\label{tab:OctupoleM1SRNS}
\end{table} 
The plots in fig.\ref{fig:OctopoleM1} give a synthetic compilation of the results showing the increase in spindown luminosity compared to flat spacetime.

For the $m=2$ octupole mode, the coupled system reads for $u^{D}_{3,2}, u^{D}_{5,2}, u^{\rm B}_{4,2}$
\begin{subequations}
\label{eq:HelmholtzOctupoleM2}
\begin{align}
 \frac{1}{r} \, \left[ r \, ( r - \Rs ) \, {u^{D}_{3,2}}'' + ( 2 \, i \, k_2 \, r^2 + \Rs ) \, {u^{D}_{3,2}}' - 12 \, u^{D}_{3,2} \right] + k_m^2 \, \left[ \left( 1 - \frac{\omega}{\Omega} \right)^2 - 1 \right] \, r \, \frac{u^{D}_{3,2}}{\alpha^2} & = -6 \, \frac{\sqrt{35}}{7} \, \varepsilon_0 \, \omega \, u^{\rm B}_{4,2} \\
 \frac{1}{r} \, \left[ r \, ( r - \Rs ) \, {u^{\rm B}_{4,2}}'' + ( 2 \, i \, k \, r^2 + \Rs ) \, {u^{\rm B}_{4,2}}' - 20 \, u^{\rm B}_{4,2} \right] + k_2^2 \, \left[ \left( 1 - \frac{\omega}{\Omega} \right)^2 - 1 \right] \, r \, \frac{u^{\rm B}_{4,2}}{\alpha^2} & = \\
 - 3 \, \mu_0 \, \omega \, \left( 2 \, \sqrt{\frac{5}{7}} \, u^{\rm D}_{3,2} - 2 \sqrt{\frac{14}{11}} \, u^{\rm D}_{5,2} \right) \\
 \frac{1}{r} \, \left[ r \, ( r - \Rs ) \, {u^{D}_{5,2}}'' + ( 2 \, i \, k_2 \, r^2 + \Rs ) \, {u^{D}_{5,2}}' - 30 \, u^{D}_{5,2} \right] + k_2^2 \, \left[ \left( 1 - \frac{\omega}{\Omega} \right)^2 - 1 \right] \, r \, \frac{u^{D}_{5,2}}{\alpha^2} & = 6 \, \sqrt{\frac{14}{11}} \, \varepsilon_0 \, \omega \, u^{\rm B}_{4,2}
\end{align}
\end{subequations}
with the boundary conditions
\begin{subequations}
\begin{align}
 \alpha^2 \, \partial_r(u_{3,2}^{\rm D}\,e^{i\,k_2\,r_*}) & = - 2 \, \sqrt{\frac{5}{7}} \, \varepsilon_0 \, r \, \tilde{\omega} \, f_{4,2}^{\rm B} \\
 \alpha^2 \, \partial_r(u_{5,2}^{\rm D}\,e^{i\,k_2\,r_*}) & = 2 \, \sqrt{\frac{14}{11}} \, \varepsilon_0 \, r \, \tilde{\omega} \, f_{4,2}^{\rm B} .
\end{align}
\end{subequations}

Integration results are shown in table~\ref{tab:OctupoleM2Schwarzschild} for Schwarzschild metric and in table~\ref{tab:OctupoleM2SRNS} for the slowly rotating neutron star metric.
\begin{table}
 \centering
\begin{tabular}{ccccccc}
\hline
\diagbox{$R/\rlight$}{$\Rs/R$} & 0 & 0.1 & 0.2 & 0.3 & 0.4 & 0.5 \\
\hline
0.01 & 0.9999 & 1.171 & 1.403 & 1.727 & 2.197 & 2.923\\
0.02 & 0.9998 & 1.179 & 1.422 & 1.760 & 2.254 & 3.017\\
0.05 & 0.9991 & 1.200 & 1.475 & 1.863 & 2.434 & 3.327\\
0.1  & 0.9967 & 1.235 & 1.567 & 2.046 & 2.770 & 3.928\\
0.2  & 0.9870 & 1.299 & 1.757 & 2.458 & 3.583 & 5.516\\
\hline
\end{tabular} 
\caption{Normalized spindown luminosity for the $m=2$ rotating octupole in Schwarzschild background.}
\label{tab:OctupoleM2Schwarzschild}
\end{table} 
\begin{table}
 \centering
\begin{tabular}{ccccccc}
\hline
\diagbox{$R/\rlight$}{$\Rs/R$} & 0 & 0.1 & 0.2 & 0.3 & 0.4 & 0.5 \\
\hline
0.01 & 0.9999 & 1.140 & 1.322 & 1.568 & 1.911 & 2.416 \\
0.02 & 0.9998 & 1.147 & 1.340 & 1.598 & 1.960 & 2.494 \\
0.05 & 0.9991 & 1.168 & 1.390 & 1.690 & 2.115 & 2.745 \\
0.1  & 0.9967 & 1.200 & 1.474 & 1.852 & 2.396 & 3.222 \\
0.2  & 0.9870 & 1.259 & 1.642 & 2.199 & 3.048 & 4.416 \\
\hline
\end{tabular} 
\caption{Normalized spindown luminosity for the $m=2$ rotating octupole in a slowly rotating metric background.}
\label{tab:OctupoleM2SRNS}
\end{table} 
The plots in fig.\ref{fig:OctopoleM2} give a synthetic compilation of the results showing the increase in spindown luminosity compared to flat spacetime.

For the $m=3$ octupole mode, the coupled system reads for $u^{D}_{3,3}, u^{D}_{5,3}, u^{\rm B}_{4,3}$
\begin{subequations}
\label{eq:HelmholtzOctupoleM3}
\begin{align}
 \frac{1}{r} \, \left[ r \, ( r - \Rs ) \, {u^{D}_{3,3}}'' + ( 2 \, i \, k_3 \, r^2 + \Rs ) \, {u^{D}_{3,3}}' - 12 \, u^{D}_{3,3} \right] + k_m^2 \, \left[ \left( 1 - \frac{\omega}{\Omega} \right)^2 - 1 \right] \, r \, \frac{u^{D}_{3,3}}{\alpha^2} & = - \sqrt{15} \, \varepsilon_0 \, \omega \, u^{\rm B}_{4,3} \\
 \frac{1}{r} \, \left[ r \, ( r - \Rs ) \, {u^{\rm B}_{4,3}}'' + ( 2 \, i \, k_3 \, r^2 + \Rs ) \, {u^{\rm B}_{4,3}}' - 20 \, u^{\rm B}_{4,3} \right] + k_3^2 \, \left[ \left( 1 - \frac{\omega}{\Omega} \right)^2 - 1 \right] \, r \, \frac{u^{\rm B}_{4,3}}{\alpha^2} & = \\
 - 3 \, \mu_0 \, \omega \, \left( \sqrt{\frac{5}{3}} \, u^{\rm D}_{3,3} - 8 \, \sqrt{\frac{2}{33}} \, u^{\rm D}_{5,3} \right) \\
 \frac{1}{r} \, \left[ r \, ( r - \Rs ) \, {u^{D}_{5,3}}'' + ( 2 \, i \, k_3 \, r^2 + \Rs ) \, {u^{D}_{5,3}}' - 30 \, u^{D}_{5,3} \right] + k_3^2 \, \left[ \left( 1 - \frac{\omega}{\Omega} \right)^2 - 1 \right] \, r \, \frac{u^{D}_{5,3}}{\alpha^2} & = 24 \, \sqrt{\frac{2}{33}} \, \varepsilon_0 \, \omega \, u^{\rm B}_{4,3}
\end{align}
\end{subequations}
with the boundary conditions
\begin{subequations}
\begin{align}
 \alpha^2 \, \partial_r(u_{3,3}^{\rm D}\,e^{i\,k_3\,r_*}) & = - \sqrt{\frac{5}{3}} \, \varepsilon_0 \, r \, \tilde{\omega} \, f_{4,3}^{\rm B} \\
 \alpha^2 \, \partial_r(u_{5,3}^{\rm D}\,e^{i\,k_3\,r_*}) & = 8\, \sqrt{\frac{2}{33}} \, \varepsilon_0 \, r \, \tilde{\omega} \, f_{4,3}^{\rm B} .
\end{align}
\end{subequations}
Integration results are shown in table~\ref{tab:OctupoleM3Schwarzschild} for Schwarzschild metric and in table~\ref{tab:OctupoleM3SRNS} for the slowly rotating neutron star metric.
\begin{table}
 \centering
\begin{tabular}{ccccccc}
\hline
\diagbox{$R/\rlight$}{$\Rs/R$} & 0 & 0.1 & 0.2 & 0.3 & 0.4 & 0.5 \\
\hline
0.01 & 0.9999 & 1.1145 & 1.267 & 1.479 & 1.787 & 2.260\\
0.02 & 0.9996 & 1.124  & 1.291 & 1.522 & 1.857 & 2.372 \\
0.05 & 0.9976 & 1.1547 & 1.364 & 1.656 & 2.084 & 2.749 \\
0.1  & 0.9906 & 1.1996 & 1.488 & 1.901 & 2.525 & 3.532 \\
0.2  & 0.9634 & 1.2745 & 1.738 & 2.465 & 3.677 & 5.865 \\
\hline
\end{tabular} 
\caption{Normalized spindown luminosity for the $m=3$ rotating octupole in Schwarzschild background.}
\label{tab:OctupoleM3Schwarzschild}
\end{table} 
\begin{table}
 \centering
\begin{tabular}{ccccccc}
\hline
\diagbox{$R/\rlight$}{$\Rs/R$} & 0 & 0.1 & 0.2 & 0.3 & 0.4 & 0.5 \\
\hline
0.01 & 0.9999 & 1.094 & 1.216 & 1.378 & 1.604 & 1.935 \\
0.02 & 0.9996 & 1.104 & 1.239 & 1.418 & 1.667 & 2.030 \\
0.05 & 0.9976 & 1.132 & 1.307 & 1.540 & 1.866 & 2.344 \\
0.1  & 0.9906 & 1.175 & 1.421 & 1.758 & 2.241 & 2.973 \\
0.2  & 0.9634 & 1.241 & 1.638 & 2.227 & 3.148 & 4.682 \\
\hline
\end{tabular} 
\caption{Normalized spindown luminosity for the $m=3$ rotating octupole in a slowly rotating metric background.}
\label{tab:OctupoleM3SRNS}
\end{table} 
The plots in fig.\ref{fig:OctopoleM3} give a synthetic compilation of the results showing the increase in spindown luminosity compared to flat spacetime.

For the $m=4$ octupole mode, the coupled system reads for $u^{D}_{5,4}, u^{\rm B}_{4,4}$
\begin{subequations}
\label{eq:HelmholtzOctupoleM4}
\begin{align}
 \frac{1}{r} \, \left[ r \, ( r - \Rs ) \, {u^{\rm B}_{4,4}}'' + ( 2 \, i \, k_4 \, r^2 + \Rs ) \, {u^{\rm B}_{4,4}}' - 20 \, u^{\rm B}_{4,4} \right] + k_4^2 \, \left[ \left( 1 - \frac{\omega}{\Omega} \right)^2 - 1 \right] \, r \, \frac{u^{\rm B}_{4,4}}{\alpha^2} & = 6 \, \sqrt{\frac{6}{11}} \, \mu_0 \, \omega \, u^{\rm D}_{5,4} \\
 \frac{1}{r} \, \left[ r \, ( r - \Rs ) \, {u^{D}_{5,4}}'' + ( 2 \, i \, k_4 \, r^2 + \Rs ) \, {u^{D}_{5,4}}' - 30 \, u^{D}_{5,4} \right] + k_4^2 \, \left[ \left( 1 - \frac{\omega}{\Omega} \right)^2 - 1 \right] \, r \, \frac{u^{D}_{5,4}}{\alpha^2} & = 6 \, \sqrt{\frac{6}{11}} \, \varepsilon_0 \, \omega \, u^{\rm B}_{4,4}
\end{align}
\end{subequations}
with the boundary conditions
\begin{subequations}
\begin{align}
 \alpha^2 \, \partial_r(u_{5,4}^{\rm D}\,e^{i\,k_4\,r_*}) & = 2 \, \sqrt{\frac{6}{11}} \, \varepsilon_0 \, r \, \tilde{\omega} \, f_{4,4}^{\rm B} .
\end{align}
\end{subequations}
Integration results are shown in table~\ref{tab:OctupoleM4Schwarzschild} for Schwarzschild metric and in table~\ref{tab:OctupoleM4SRNS} for the slowly rotating neutron star metric.
\begin{table}
 \centering
\begin{tabular}{ccccccc}
\hline
\diagbox{$R/\rlight$}{$\Rs/R$} & 0 & 0.1 & 0.2 & 0.3 & 0.4 & 0.5 \\
\hline
0.01 & 0.9997 & 1.012 & 1.025 & 1.038 & 1.051 & 1.065 \\
0.02 & 0.9990 & 1.024 & 1.050 & 1.077 & 1.106 & 1.135 \\
0.05 & 0.9942 & 1.057 & 1.127 & 1.202 & 1.285 & 1.377 \\
0.1  & 0.9773 & 1.104 & 1.253 & 1.429 & 1.641 & 1.898 \\
0.2  & 0.9118 & 1.152 & 1.478 & 1.929 & 2.572 & 3.520 \\
0.5  & 0.5417 & 0.891 & 1.548 & 2.887 & 5.906 & 13.75 \\
\hline
\end{tabular} 
\caption{Normalized spindown luminosity for the $m=4$ rotating octupole in Schwarzschild background.}
\label{tab:OctupoleM4Schwarzschild}
\end{table} 
\begin{table}
 \centering
\begin{tabular}{ccccccc}
\hline
\diagbox{$R/\rlight$}{$\Rs/R$} & 0 & 0.1 & 0.2 & 0.3 & 0.4 & 0.5 \\
\hline
0.01 & 0.9997 & 1.012 & 1.025 & 1.038 & 1.051 & 1.065 \\
0.02 & 0.9990 & 1.024 & 1.050 & 1.077 & 1.105 & 1.134 \\
0.05 & 0.9942 & 1.057 & 1.125 & 1.199 & 1.281 & 1.370 \\
0.1  & 0.9773 & 1.101 & 1.245 & 1.415 & 1.617 & 1.860 \\
0.2  & 0.9118 & 1.140 & 1.444 & 1.856 & 2.427 & 3.241 \\
\hline
\end{tabular} 
\caption{Normalized spindown luminosity for the $m=4$ rotating octupole in a slowly rotating metric background.}
\label{tab:OctupoleM4SRNS}
\end{table} 
The plots in fig.\ref{fig:OctopoleM4} give a synthetic compilation of the results showing the increase in spindown luminosity compared to flat spacetime.

\begin{figure}
\centering
\input{spindown_octopole_m1_NR30.tex}
\caption{Spindown luminosity of the orthogonal rotator ocotpole for the mode $m=1$ in Schwarzschild and slowly rotating neutron star metric approximation.}
\label{fig:OctopoleM1}
\end{figure}

\begin{figure}
\centering
\input{spindown_octopole_m2_NR30.tex}
\caption{Spindown luminosity of the orthogonal rotator ocotpole for the mode $m=2$ in Schwarzschild and slowly rotating neutron star metric approximation.}
\label{fig:OctopoleM2}
\end{figure}

\begin{figure}
\centering
\input{spindown_octopole_m3_NR30.tex}
\caption{Spindown luminosity of the orthogonal rotator ocotpole for the mode $m=3$ in Schwarzschild and slowly rotating neutron star metric approximation.}
\label{fig:OctopoleM3}
\end{figure}

\begin{figure}
\centering
\input{spindown_octopole_m4_NR30.tex}
\caption{Spindown luminosity of the orthogonal rotator ocotpole for the mode $m=4$ in Schwarzschild and slowly rotating neutron star metric approximation.}
\label{fig:OctopoleM4}
\end{figure}

\subsection{Contributions to spindown luminosity}

\cor{As we saw in the Newtonian case, the highest contribution to the luminosity does not necessarily correspond to the magnetic multipole of order $(\ell,m)$ supposed to be anchored in the neutron star. We showed thanks to exact analytical solutions expressed in terms of spherical Hankel functions that for $\ell>m$ it is always the electric multipole of order $(\ell-1,m)$ that contributes most to the energy loss. This trend should also be observed in the general-relativistic case. Thus for some special configurations of magnetic field, rotation and gravity, we computed separately the contribution from the magnetic multipole and the electric multipole to check the relevance of both part to the total Poynting flux. The results are summarized in the table~\ref{tab:LuminositeContribution}.
For the magnetic dipole $(\ell,m)=(1,1)$, the dipolar component radiates very efficiently, it represents almost 100\% of the energy losses, the electric quadrupolar part is irrelevant, contributing to one part to a million. Adding space curvature and frame dragging effects does not change our conclusion. A very good estimate of the spindown requires only the knowledge of the magnetic radiating part.
For the magnetic quadrupole $(\ell,m)=(2,1)$, the situation is quite opposite. Indeed the quadrupolar magnetic component radiates only about 15\% of the total losses, it represents a negligible part of the dynamics. However, the electric dipolar part is highest, contributing to about 85\%. Adding space curvature and frame dragging effects does not drastically change the magnetic contribution but the electric part is enhanced by a factor around 3.5. In this configuration, a very good estimate of the spindown requires only the knowledge of the electric radiating part. Frame dragging decreases slightly the flux.
For the magnetic quadrupole $(\ell,m)=(2,2)$, we retrive the behavior reminescent of the magnetic dipole. Indeed the quadrupolar magnetic component radiates almost exclusively whereas the electric hexapolar part completely negligible. Space curvature and frame dragging effects enhance the magnetic radiating part slightly.
Going briefly through the hexapolar and octopolar magnetic field, the previous findings apply in the same way. For the mode $\ell=m$, the magnetic part is highest and the electric part remains negligible. General relativity slightly increases the luminosity of this magnetic multipole radiation. For all other cases with $\ell>m$, the electric part radiates at the highest level and the magnetic part just adds some corrections. However, general relativity enhances the losses from the electric multipole, multipliyng the rate by a factor of several units. A corollary of this effect is a drastic difference between flat spacetime and Schwarzschild spacteime Poynting flux for point multipoles. General relativistic expectations do not converge to Newtonian energy losses when $R/\rlight\to0$. To my knowledge this subtle non cancellation has never been noticed in the literature.}

\begin{table}
 \centering
\begin{tabular}{c|cc|cc|cc}
\hline
Multipole & \multicolumn{2}{c}{flat} & \multicolumn{2}{c}{Schw} & \multicolumn{2}{c|}{SRNS} \\
\cline{2-7}
    & B-$(\ell,m)$ & E-$(\ell\pm1,m)$ & B-$(\ell,m)$ & E-$(\ell\pm1,m)$ & B-$(\ell,m)$ & E-$(\ell\pm1,m)$ \\
\hline
\hline
(1,1) & 0.9901 & 1.667e-6 & 1.166 & 1.505e-6 & 1.163 & 1.874e-6 \\
(2,1) & 0.1557 & 0.8521 & 0.1832 & 3.031 & 0.1832 & 2.524 \\
(2,2) & 0.9866 & 2.026e-6 & 1.372 & 2.414e-6 & 1.361 & 2.948e-6 \\
(3,1) & 0.08029 & 0.9195 & 0.09436 & 3.457 & 0.09434 & 2.834 \\
(3,2) & 0.3561 & 0.6411 & 0.4942 & 2.862 & 0.492 & 2.324 \\
(3,3) & 0.9821 & 1.901e-6 & 1.614 & 2.827e-6 & 1.591 & 3.406e-6 \\
(4,1) & 0.04884 & 0.9504 & 0.05736 & 3.691 & 0.05745 & 2.996 \\
(4,2) & 0.2033 & 0.7933 & 0.2817 & 3.647 & 0.2810 & 2.941 \\
(4,3) & 0.4915 & 0.4990 & 0.8056 & 2.726 & 0.7976 & 2.176 \\
(4,4) & 0.9773 & 1.706e-6 & 1.898 & 3.097e-6 & 1.860 & 3.692e-6\\
\hline
\end{tabular} 
\caption{Normalized spindown contribution from the magnetic multipole~B-$(\ell,m)$ and electric multipole~E-$(\ell\pm1,m)$ for a rotating multipole of order~$(\ell,m)$ in the flat, Schwarzschild (Schw) and slowly rotating (SRNS) metric. The sign of E-$(\ell\pm1,m)$ is choosen such that $\ell\pm1\geqslant m$ for the lowest possible $\ell$. The plus sign applies only when $\ell=m$. The parameter used are $R/\rlight=0.1$ and $\Rs/R=0.5$ (if relevant).}
\label{tab:LuminositeContribution}
\end{table}

\subsection{Fitting expressions}

As a summary of all the above results, we give fitting expressions for the spindown luminosity for each multipole. Only for the modes $\ell=m$ the spindown luminosity tends to the flat spacetime value when $R/\rlight\rightarrow0$. For the other modes, it seems that the Poynting flux stay larger than its flat spacetime counterpart even for a point multipole. 

The approximate formula used to numerically fit the data for the luminosity increase includes terms up to quadratic order such that the model function is
\begin{equation}
\frac{L_{\rm GR}}{L_{\rm flat}} = \mathcal{S}(a,\Xi) = 1 + \Xi \, ( l_1 + l_2 \, a + l_3 \, a^2 ) + \Xi^2 \, ( l_4 + l_5 \, a + l_6 \, a^2 ) .
\end{equation}
The choice of this formal expression is dictated by the fact that for zero compactness~$\Xi=0$ the spindown luminosity converges to the flat spacetime limit thus $\frac{L_{\rm GR}}{L_{\rm flat}}=1$. We then added corrections to first and second order in compactness~$\Xi$ therefore also keeping second order corrections in the spin parameter~$a$. The accuracy of these fits dependent strongly on the multipole we try to adjust. Best results are obtained for the dipole. Nevertheless, all fits give accuracy better than 20\%. The list of coefficients and the maximal error in the fit are summarized in table~\ref{tab:Fit}.
\begin{table}
 \centering
\begin{tabular}{cccccccc}
\hline
Multipole $(\ell,m)$ & $l_1$ & $l_2$ & $l_3$ & $l_4$ & $l_5$ & $l_6$ & max(error) \\
\hline
(1,1) & -0.00114 & 3.206 & -2.989 & -0.0009536 & 0.1251 & 10.69 & 0.001606 \\
(2,1) & 0.7208 & 1.706 & -5.975 & 3.627 & 11.08 & 23.95 & 0.05797 \\
(2,2) & -0.002069 & 6.542 & -8.615 & 0.005343 & -0.4638 & 42.52 & 0.007425 \\
(3,1) & 0.7691 & 1.517 & -5.751 & 4.359 & 12.14 & 28.33 & 0.06858 \\
(3,2) & 0.5325 & 4.368 & -20.76 & 3.041 & 16.33 & 97.29 & 0.09678. \\
(3,3) & -0.007845 & 10.3 & -20.15 & 0.02878 & -2.698 & 108.1 & 0.02187 \\
(4,1) & 0.7479 & 2.176 & -8.628 & 4.641 & 15.48 & 18.94 & 0.07213 \\
(4,2) & 0.6308 & 4.201 & -24.79 & 3.98 & 20.82 & 111.6 & 0.1204 \\
(4,3) & 0.382  & 8.505 & -47.93 & 2.564 & 15.63 & 225.2 & 0.1508 \\
(4,4) & -0.03599 & 15.14 & -42.8 & 0.08634 & -8.143 & 226.3 & 0.05161 \\
\hline
\end{tabular} 
\caption{Best fit of the spindown luminosity for all multipoles, frame dragging included.}
\label{tab:Fit}
\end{table} 
We notice that for the cases $\ell=m$ the constant $l_1$ and $l_4$ almost vanish. This means that in the limit $a\rightarrow0$ that is for very slow rotators, the general relativistic luminosity tends to the flat spacetime limit. This conclusion does not hold for the other multipoles satisfying $\ell\neq m$. There is always a larger spindown in general relativity compared to Newtonian gravity.

\subsection{Braking index}

As a diagnostic of multipole fields, we computed the Poynting flux of single multipoles labelled by the mode $(\ell,m)$ taking into account the finite size of the star, space curvature and frame dragging. The braking index~$n$ is another interesting related quantity which describes the efficiency of electromagnetic braking by the law $\dot{\Omega} = - K \, \Omega^n$ where $K$ is a constant depending on boundary conditions on the neutron star. For magnetic multipolar point sources of order~$\ell$, we know that $n_\ell=2\,\ell+1$ \citep{krolik_multipolar_1991, petri_multipolar_2015}. It is well known for instance that the braking index for magnetodipole losses is equal to $n=3$ within small corrections in power of $R/\rlight$. However in general the braking index can differ from this fiducial value if the size of the star is taken into account. What about corrections due to general relativity? To answer this question, we compute the braking index derived from the spindown luminosity expressed in terms of the dimensionless parameter~$a=R/\rlight$ according to
\begin{equation}
\label{eq:IndiceFreinage}
 n_{\rm GR} = \frac{a}{L_{\rm GR}} \, \frac{dL_{\rm GR}}{da} - 1 = \frac{d\ln L_{\rm GR}}{d\ln a} - 1.
\end{equation}
This is the general formula to compute the braking index in any case, knowing the luminosity of the star with respect to the spin normalized by the parameter~$a$. Taking into account the general-relativistic correction factor compared to flat spacetime, we get
\begin{equation}
\label{eq:IndiceFreinageCorrection}
 n_{\rm GR} = \frac{d\ln \mathcal{S}}{d\ln a} + \frac{d\ln L_{\rm flat}}{d\ln a} - 1 = n_{\rm flat} + \frac{d\ln \mathcal{S}}{d\ln a} 
\end{equation}
where $n_{\rm flat}$ is the barking index obtained from flat spacetime multipoles, the one computed in \cite{petri_multipolar_2015}. General-relativity clearly introduces some corrections to the standard value expected from Newtonian gravity. From the values of the fitting coefficients, numerical applications show that the variation in the braking index remains bounded to small increase not higher than 0.3-0.4. General-relativistic effects cannot account for large discrepancies between measured braking indexes of pulsars and the fiducial value of the point dipole equal to $n=3$.

\section{TIME-DEPENDENT SIMULATIONS}
\label{sec:Simulations}

\cor{As a final consistency check of our results we performed some time-dependent numerical simulations of the vacuum Maxwell equations in general relativity for a magnetic dipole with $(\ell,m)=(1,1)$, a magnetic quadrupole with $(\ell,m)=(2,1)$ and $(\ell,m)=(2,2)$ and a magnetic hexapolar with  $(\ell,m)=(3,1)$, $(\ell,m)=(3,2)$ and $(\ell,m)=(3,2)$. The relevant parameters of the simulations are a spin rate of $R/\rlight=0.1$ and use of the three metrics: Minkowski ($\Rs/R=0$), Schwarzschild ($\Rs/R=0.5$) and slowly rotating neutron star ($\Rs/R=0.5$). The pseudo-spectral discontinuous Galerkin technique using he 3+1 formalism of general relativity has been exposed in depth in \cite{petri_general-relativistic_2013} and is not reproduced here. }

\cor{A summary of our results are given in table~\ref{tab:LuminositeSimulation}. These values are in perfect agreement with our semi-analytical treatment resolving the stationary problem with Helmholtz equations and expansion into rational Chebyshev functions of the generalized spherical Hankel functions. Although the numerical simulations take into account many magnetic multipoles due to coupling by frame dragging, in practice only the lowest magnetic multipole is required for a good approximate estimate.
\begin{table}
 \centering
\begin{tabular}{cccc}
\hline
Multipole & flat & Schw & SRNS \\
\hline
\hline
(1,1) & 0.987 & 1.163 & 1.160 \\
(2,1) & 1.006 & 3.209 & 2.704 \\
(2,2) & 0.987 & 1.373 & 1.362 \\
(3,1) & 0.995 & 3.583 & 3.022 \\
(3,2) & 0.997 & 3.446 & 2.906 \\
(3,3) & 0.982 & 1.580 & 1.558 \\
\hline
\end{tabular} 
\caption{Normalized spindown extracted from the time-dependent numerical simulations for $R/\rlight=0.1$ and $\Rs/R=0.5$ (if applicable).}
\label{tab:LuminositeSimulation}
\end{table} 
}

\cor{The good agreement gives us confidence about the consistency of our analytical, semi-analytical and numerical approaches to investigate some properties of general-relativistic rotating magnetic multipoles.}

\section{CONCLUSION}
\label{sec:Conclusion}

Computing general-relativistic extensions of the Deutsch solution including space curvature and frame-dragging effects is easily achieved numerically to very high precision thanks to appropriate change of the coordinate system and a judicious expansion of the unknown functions onto rational Chebyshev functions. These spectral methods even allow to get reasonable approximate analytical solutions to the Helmholtz equations in the slow rotation metric approximation. These expressions are very handy to deduce important properties of rotating multipoles in general relativity. In this spirit, we estimated the corrections to the electromagnetic spindown luminosity as well as to the braking index of pulsar in the parameter range of compactness and periods relevant to neutron star electrodynamics. General-relativity increases the spindown luminosity with respect to Newtonian gravity although frame-dragging does not significantly impact on this estimates. Also for realistic pulsar parameters the braking index remains essentially unaffected by gravity perturbations.

Future time-dependent numerical simulations of rotating multipoles in general relativity should confirm our estimates given in this paper. As an example of multipolar electromagnetic field, we plane to extend our method to a rotating off-centred dipole in general relativity. Investigation of pulse profiles and related polarisation characteristics will offer a valuable insight into the magnetic topology close to the neutron star surface in the strong gravity regime. \cor{Last but not least, the effect of plasma charges and currents circulating inside the magnetosphere must be carefully included into the picture of magnetic multipoles to point out the effect of frame dragging which will become sensitive to first order in spin frequency.}

\section*{Acknowledgements}

This work has been supported by the French National Research Agency (ANR) through the grant No. ANR-13-JS05-0003-01 (project EMPERE). \cor{It also benefited from the computational facilities available at Equip@Meso from Universit\'e de Strasbourg.}


\label{lastpage}

\end{document}